\documentclass[aps, prd, superscriptaddress, preprintnumbers, showpacs]{revtex4}

\usepackage{graphicx,amsfonts,amsmath,amssymb,amstext}
\usepackage{float,wrapfig}
\usepackage{subfigure, psfrag}
\usepackage{dsfont}

\usepackage{color}
\usepackage{bm}

\newcommand{\be}{\begin{equation}}
\newcommand{\ee}{\end{equation}}
\newcommand{\ba}{\begin{eqnarray}}
\newcommand{\ea}{\end{eqnarray}}

\definecolor{purple}{rgb}{0.8,0,0.6}

\begin{document}

\title{Surface Fermi arcs in $\mathbb{Z}_2$ Weyl semimetals $\mathbf{A_3Bi}$
($\mathbf{A}=\mathbf{Na}$, $\mathbf{K}$, $\mathbf{Rb}$)}
\date{\today}

\author{E. V. Gorbar}
%\email{gorbar@bitp.kiev.ua}
\affiliation{Department of Physics, Taras Shevchenko National Kiev University, Kiev, 03680, Ukraine}
\affiliation{Bogolyubov Institute for Theoretical Physics, Kiev, 03680, Ukraine}

\author{V. A. Miransky}
%\email{vmiransk@uwo.ca}
\affiliation{Department of Applied Mathematics, Western University, London, Ontario N6A 5B7, Canada}
\affiliation{Department of Physics and Astronomy, Western University, London, Ontario N6A 3K7, Canada}

\author{I. A. Shovkovy}
%\email{igor.shovkovy@asu.edu}
\affiliation{College of Letters and Sciences, Arizona State University, Mesa, Arizona 85212, USA}

\author{P. O. Sukhachov}
%\email{pavel.sukhachov@gmail.com}
\affiliation{Department of Physics, Taras Shevchenko National Kiev University, Kiev, 03680, Ukraine}

\begin{abstract}
The surface Fermi arc states in $\mathbb{Z}_2$ Weyl semimetals $\mathrm{A_3Bi}$
($\mathrm{A}=\mathrm{Na}$, $\mathrm{K}$, $\mathrm{Rb}$) are studied by employing
a continuum low-energy effective model. It is shown that the surface Fermi arc states can
be classified with respect to the ud-parity symmetry. Because of the symmetry, the arcs come
in mirror symmetric pairs. The effects of symmetry breaking terms on the structure of the
Fermi arcs are also studied. Among other results, we find at least two qualitatively different
types of the surface Fermi arcs. The arcs of the first type link {\it disconnected} sheets of
the bulk Fermi surface, while arcs of the second type link different points of the {\it same}
bulk Fermi surface sheet.
\end{abstract}

%71.70.Di	Landau levels

\maketitle

\section{Introduction}
\label{sec:Introduction}

3D Dirac semimetals are 3D analogs of graphene \cite{Geim}. Their conduction
and valence bands touch only at discrete (Dirac) points in the Brillouin zone with
the electron states described by the 3D massless Dirac equation.
Each Dirac point in momentum space is composed of two superimposed
Weyl nodes of opposite chirality. Such points are usually obtained by fine tuning of
certain physical parameters (e.g., the spin-orbit coupling strength or chemical
composition) and are difficult to control. Additionally, they are often unstable with
respect to mixing of Weyl modes and opening a gap.

An important idea was proposed in Refs.~\cite{Manes:2011jk,Mele}, where
it was shown that an appropriate crystal symmetry can protect and
stabilize the gapless 3D Dirac points. Indeed, if a pair of crossing bands
belong to different irreducible representations of the discrete (rotational) crystal
symmetry and if this symmetry is not broken dynamically,
then the mass term for the corresponding Dirac fermions will be prohibited.
The {\it ab initio} calculations in Ref.~\cite{Mele} showed that $\beta$-cristobalite
$\mathrm{BiO_2}$ exhibits three Dirac points at the Fermi level.
Unfortunately, this material is metastable. By using the first-principles calculations
and an effective model analysis, the compounds $\mathrm{A_3Bi}$ (A=Na, K, Rb)
and $\mathrm{Cd_3As_2}$ were identified in Refs.~\cite{Fang,WangWeng} as
possible 3D Dirac semimetals protected by crystal symmetry. Giant diamagnetism,
linear quantum magnetoresistance, and the quantum spin Hall effect are
expected in these materials. Furthermore, various topologically distinct
phases can be realized in these compounds by breaking the time-reversal and
inversion symmetries. By using angle-resolved photoemission spectroscopy,
the Dirac semimetal band structure was indeed observed \cite{Borisenko,Neupane,Liu}
in $\mathrm{Cd_3As_2}$ and $\mathrm{Na_3Bi}$ opening the path toward experimental
investigations of the properties of 3D Dirac semimetals.

Weyl semimetals is another group of materials that is closely related to 3D Dirac semimetals
and have already attracted a lot of theoretical interest (for reviews, see Refs.~\cite{Hook,Turner,Vafek}).
They are characterized by topologically non-trivial Weyl nodes in reciprocal space. Weyl nodes
are monopoles of the Berry flux and, therefore, can appear or annihilate only in pairs. Weyl
semimetals were proposed to be realized in pyrochlore iridates \cite{Savrasov}, topological
heterostructures \cite{Balents}, magnetically doped topological insulators \cite{Cho}, and
nonmagnetic materials such as $\mathrm{TaAs}$ \cite{1501.00060,1501.00755}. Recently,
first experimental studies of Weyl semimetal candidate $\mathrm{TaAs}$ were reported in
Refs.~\cite{1502.00251,1502.03807,1502.04684,1503.01304}. The authors observed
unusual transport properties and surface states that are characteristic of the Weyl
semimetal phase. Another interesting realization of the Weyl points in the context of 
photonic crystals has been recently reported in Ref.~\cite{Weyl-photonic}.

Since the magnetic field breaks the time-reversal symmetry, a Dirac (semi-)metal in a
magnetic field may transform into a Weyl one with Weyl nodes separated in
momentum space by a nonzero chiral shift \cite{Gorbar:2013qsa}.
Experimentally, the transition from a Dirac metal to a Weyl one in
a magnetic field might have been observed in $\mathrm{Bi_{1-x}Sb_x}$ for
$x \approx 0.03$ \cite{Kim:2013dia}. In moderately strong magnetic fields,
a negative magnetoresistivity is observed and interpreted as a fingerprint
\cite{Nielsen:1983rb,Son:2012bg,Gorbar:2013dha} of a Weyl/Dirac metal phase.

The surface Fermi arcs \cite{Savrasov,Haldane,Aji,Okugawa:2014ina}, which
connect Weyl nodes of opposite chirality, are related to the non-trivial topology
of Weyl semimetals. In equilibrium, the presence of such surface
states ensures that the chemical potentials at different Weyl points are identical
\cite{Haldane}. Although Fermi arcs always connect Weyl nodes of opposite
chirality, their shapes depend on the boundary conditions and, as
shown in Ref.~\cite{Hosur}, Fermi arcs of an arbitrary form can be engineered.
The Fermi arcs on the opposite surfaces of a semimetal sample together with
the Fermi surfaces of bulk states form a closed Fermi surface. In an
external magnetic field, the nontrivial structure of the corresponding Fermi surface
gives rise to closed magnetic orbits involving the surface Fermi arcs \cite{Vishwanath}.
These orbits produce periodic quantum oscillations of the density of
states in a magnetic field leading to unconventional Fermiology of surface states.
It was argued in Ref.~\cite{Gorbar:2014qta} that the interaction effects
can change the separation between Weyl nodes in momentum space
and the length of the Fermi arcs in the reciprocal space and, thus, affect these
magnetic orbits. As a result, we found that the period of oscillations of
the density of states related to closed magnetic orbits involving
Fermi arcs has a non-trivial dependence on the orientation of the
magnetic field projection in the plane of the semimetal surface \cite{Gorbar:2014qta}.
If experimentally observed, such a dependence would provide an important
clue to the effects of interactions in Weyl semimetals.

Normally, one would not expect any surface Fermi arcs in 3D Dirac semimetals
because the Dirac point has no topological charge and the associated Berry flux vanishes.
In Refs.~\cite{Fang,WangWeng}, however, it was shown that the 3D Dirac
semimetals $\mathrm{A_3Bi}$ (A=Na, K, Rb) and $\mathrm{Cd_3As_2}$ possess
non-trivial surface Fermi arcs. This finding suggests a topologically nontrivial nature
of the corresponding Dirac materials. Recently we showed \cite{Gorbar:2014sja} that this is indeed the
case for Dirac semimetals $\mathrm{A_3Bi}$ ($\mathrm{A}=\mathrm{Na}, \mathrm{K}, \mathrm{Rb}$).
The physical reason for their nontrivial topological properties is connected with a discrete
symmetry of the low-energy effective Hamiltonian. The symmetry classification allows one
to split all electron states into two separate sectors, each describing a Weyl semimetal with
a pair of Weyl nodes and broken time-reversal symmetry. The time-reversal symmetry is
preserved in the complete theory because its transformation interchanges states from the
two different sectors. The nontrivial topological structure of each sector was supported by explicit calculations
of the Berry curvature, which revealed a pair of monopoles of the Berry flux at the positions of
Weyl nodes in each of the two sectors of these semimetals \cite{Gorbar:2014sja}. In essence,
these results demonstrated that Dirac semimetals $\mathrm{A_3Bi}$ ($\mathrm{A}=\mathrm{Na},
\mathrm{K}, \mathrm{Rb}$) are, in fact, $\mathbb{Z}_2$ Weyl semimetals.

In Refs.~\cite{Fang,WangWeng}, the surface Fermi arcs in 3D Dirac semimetals were obtained
in a tight-binding model by using an iterative method that produces the surface Green's function
of the semi-infinite system \cite{Yu}. The imaginary part of the surface Green's function makes
possible to determine the local density of states at the surface. While such a technique is very
powerful, it is essentially a ``black box". In contrast, in the present paper, we study
analytically the surface Fermi arc states by employing the continuum low-energy effective model
with appropriate boundary conditions at the surface. We hope that such a consideration will
provide a deeper understanding of the physical properties and characteristics of the surface Fermi arcs,
as well as shed more light on the nontrivial topological properties of the $\mathrm{A_3Bi}$ compounds.

The paper is organized as follows. In Sec.~\ref{sec:model}, we introduce the low-energy effective
model and discuss its symmetries. The recently revealed $\mathbb{Z}_2$ Weyl semimetal
structure of $\mathrm{A_3Bi}$ ($\mathrm{A}=\mathrm{Na}$, $\mathrm{K}$, $\mathrm{Rb}$)
is emphasized. In order to clarify the origin and the structure of the surface Fermi arcs,
we study in Sec.~\ref{sec:2x2Model} the corresponding states in a simplified model
that contains a single Weyl semimetal sector. In Sec.~\ref{sec:Realistic4x4Model}, we present the
rigorous analysis of the surface Fermi arc states in a realistic low-energy model of
semimetals $\mathrm{A_3Bi}$ ($\mathrm{A}=\mathrm{Na}$, $\mathrm{K}$, $\mathrm{Rb}$).
The effects of several possible symmetry breaking terms on the structure of the surface Fermi
arc states are investigated in Sec.~\ref{sec:DynamicalParametersModel}. The
discussion and the summary of the main results are given in Sec.~\ref{Conclusion}.
Technical details regarding the symmetry properties and classification of the Fermi arc
states are presented in Appendices~\ref{AppExtra} and \ref{AppA}.

For convenience, throughout the paper, we set $\hbar=1$ and $c=1$.

\section{Model}
\label{sec:model}

\subsection{Low-energy effective Hamiltonian}
\label{effective-Hamiltonian}

The low-energy Hamiltonian derived in Ref.~\cite{Fang} for $\mathrm{A_3Bi}$
($\mathrm{A}=\mathrm{Na}, \mathrm{K}, \mathrm{Rb}$) has the form
\begin{equation}
H(\mathbf{k}) = \epsilon_0(\mathbf{k}) + H_{4\times 4},
\label{low-energy-Hamiltonian}
\end{equation}
where $\epsilon_0(\mathbf{k}) = C_0 + C_1k_z^2+C_2(k_x^2+k_y^2)$ and
\begin{equation}
H_{4\times 4} =
\left( \begin{array}{cccc}
                   M(\mathbf{k}) & Ak_+ & 0 & B^{*}(\mathbf{k}) \\
                   Ak_- & -M(\mathbf{k}) & B^{*}(\mathbf{k}) & 0 \\
                   0 & B(\mathbf{k}) & M(\mathbf{k}) & -Ak_- \\
                   B(\mathbf{k}) & 0 & -Ak_+ & -M(\mathbf{k}) \\
        \end{array}
\right).
\label{low-energy-Hamiltonian4x4}
\end{equation}
While the diagonal elements of $H_{4\times 4}$ are given in terms of a single function,
$M(\mathbf{k}) = M_0 - M_1 k_z^2-M_2(k_x^2+k_y^2)$, the off-diagonal elements
are determined by functions $Ak_{\pm}$ and $B(\mathbf{k}) = \alpha k_zk_{+}^2$,
where $k_{\pm} = k_x\pm ik_y$.

By fitting the energy spectrum of the effective Hamiltonian with the {\em ab initio}
calculations, the numerical values of parameters in the effective model were
determined in Ref.~\cite{Fang}. They are
\begin{equation}
\begin{array}{lll}
 C_0 = -0.06382~\mbox{eV},\qquad
& C_1 = 8.7536~\mbox{eV\,\AA}^2,\qquad
& C_2 = -8.4008~\mbox{eV\,\AA}^2,\\
 M_0=-0.08686~\mbox{eV},\quad
& M_1=-10.6424~\mbox{eV\,\AA}^2,\qquad
& M_2=-10.3610~\mbox{eV\,\AA}^2,\\
 A=2.4598~\mbox{eV\,\AA},\qquad
& a=5.448~\mbox{\AA},\qquad
& c=9.655~\mbox{\AA},
\end{array}
\label{model-parameters}
\end{equation}
where we also included the lattice constants $a$ and $c$. Since no specific value for $\alpha$
was quoted in Ref.~\cite{Fang}, we will treat it as a free parameter below.

The energy eigenvalues of the low-energy Hamiltonian (\ref{low-energy-Hamiltonian})
are given by the following explicit expression:
\begin{equation}
E(\mathbf{k})=\epsilon_0(\mathbf{k}) \pm \sqrt{M^2(\mathbf{k})+A^2k_{+}k_{-}+|B(\mathbf{k})|^2}.
\label{energy-dispersion}
\end{equation}
It is easy to check that the term with the square root vanishes at the two Dirac points,
$\mathbf{k}^{\pm}_0=\left(0, 0,  \pm \sqrt{m}\right)$, where $\sqrt{m}\equiv \sqrt{M_0/M_1}$.
With the choice of the low-energy parameters in Eq.~(\ref{model-parameters}), we find that
$\sqrt{m}\approx 0.09034~\mbox{\AA}^{-1}$. The function $B(\mathbf{k})$ plays the role of
a momentum dependent mass (gap) function that vanishes at the Dirac points.

It is instructive to show that linearizing $M(\mathbf{k})$ in the vicinity of the Dirac points $\mathbf{k}^{\pm}_0$,
Hamiltonian (\ref{low-energy-Hamiltonian4x4}) takes the form of a 3D massive Dirac
Hamiltonian. In the vicinity of $\mathbf{k}^{-}_0$, expanding $M(\mathbf{k})$ to the linear order in
$\mathbf{\delta{k}}=\mathbf{k}-\mathbf{k}^{-}_0$, we obtain
\begin{equation}
H^{\rm lin}_{4\times 4}=\left(
         \begin{array}{cc}
          A(\tilde{k}_x\sigma_x-\tilde{k}_y\sigma_y-\tilde{k}_z\sigma_z)  & B^{*}(\mathbf{k}) \sigma_x \\
           B(\mathbf{k}) \sigma_x & -A\,\mathbf{\tilde{k}}\cdot\bm{\sigma} \\
         \end{array}
       \right),
\label{model-Hamiltonian-1}
\end{equation}
where $\bm{\sigma}$ are Pauli matrices and $\mathbf{\tilde{k}}=(k_x,k_y,2\delta k_z\sqrt{M_0M_1}/A)$.
Furthermore, by performing the unitary transformation, $\tilde{H}^{\rm lin}_{4\times 4}\equiv U_x^{+}H^{\rm lin}_{4\times 4}U_x$,
where $U_x=\mbox{diag}( \sigma_x , I_2)$ and $I_2$ is the $2\times 2$ unit matrix, we find that the Hamiltonian takes the standard
form of the Dirac Hamiltonian in the chiral representation,
\begin{equation}
\tilde{H}^{\rm lin}_{4\times 4}=\left(
         \begin{array}{cc}
          A\,\mathbf{\tilde{k}}\cdot\bm{\sigma} & B^{*}(\mathbf{k}) \\
          B(\mathbf{k}) & -A\,\mathbf{\tilde{k}}\cdot\bm{\sigma} \\
         \end{array}
       \right).
\label{model-Hamiltonian-canonical}
\end{equation}
Taking into account that the mass term $B(\mathbf{k})$ vanishes at the Dirac point, we conclude that the
upper and lower $2 \times 2$ blocks describe quasiparticle states of opposite chiralities. Also, since
the leading order nonzero corrections to the mass function are quadratic in momentum, the
chirality remains a good quantum number in a sufficiently small vicinity of the Dirac point.
Hamiltonian (\ref{model-Hamiltonian-canonical}), describing two subsets of the
opposite chirality states near a single Dirac point, does not appear to have any interesting
topological properties. Also, by itself, it is unlikely to give rise to any Fermi arcs states. It is easy to
check, however, that Hamiltonian (\ref{low-energy-Hamiltonian4x4}) linearized near $\mathbf{k}^{+}_0$
has a similar structure and describes two additional subsets of the opposite chirality states. As we
argue below, the superposition of the two sectors of the theory is nontrivial and gives rise
to an interesting topological structure \cite{Gorbar:2014sja}.

\subsection{Symmetries}
\label{sec:symmetries}

Let us briefly review the symmetry properties of the low-energy Hamiltonian following
Ref.~\cite{Gorbar:2014sja}. We start by pointing out that, as expected, the Hamiltonian
(\ref{low-energy-Hamiltonian}) is invariant under the time-reversal and inversion
symmetries, i.e.,
\begin{eqnarray}
\Theta H_{-\mathbf{k}} \Theta^{-1} &=& H_{\mathbf{k}}, \qquad \mbox{(time-reversal symmetry)}
\label{T-symmetry}
\\
PH_{\mathbf{-k}}P^{-1} &=& H_{\mathbf{k}},\qquad \mbox{(inversion symmetry)}
\label{P-symmetry}
\end{eqnarray}
where $\Theta=TK$ ($K$ is complex conjugation) and
\begin{equation}
T = \left(
       \begin{array}{cccc}
           0 & 0 & 1 & 0 \\
           0 & 0 & 0 & 1  \\
           -1 & 0 & 0 & 0 \\
           0 & -1 & 0 & 0
         \end{array}
       \right),
\qquad
P=\left(
       \begin{array}{cccc}
           1 & 0 & 0 & 0 \\
           0 & -1 & 0 & 0  \\
           0 & 0 & 1 & 0 \\
           0 & 0 & 0 & -1
         \end{array}
       \right).
\label{P-matrix}
\end{equation}
Having both, the time-reversal and inversion symmetries, suggests that the corresponding
compounds are not Weyl semimetals. This is not the whole story, however.

As shown in Ref.~\cite{Gorbar:2014sja}, the low-energy Hamiltonian in Eq.~(\ref{low-energy-Hamiltonian})
possesses a new discrete symmetry, the so-called up-down parity (ud-parity), that protects its topological
nature. In order to understand the corresponding symmetry, it is instructive to start from the approximate
Hamiltonian without the mass function $B(\mathbf{k})$ (or, alternatively, $\alpha =0$). In this case, the
$4\times 4$ Hamiltonian takes a block diagonal form: 
$H_{4\times 4}(\alpha=0) \equiv H^{+}_{2\times 2}\oplus H^{-}_{2\times 2}$.
The explicit form of the upper block is given by
\begin{eqnarray}
\label{Hamiltonians_+_New}
&&H^{+}_{2\times 2}= \left(
         \begin{array}{cc}
           M_0-M_1k^2_z-M_2(k^2_x+k^2_y) & A(k_x+ik_y) \\
           A(k_x-ik_y) & -\left[M_0-M_1k^2_z-M_2(k^2_x+k^2_y)\right]  \\
         \end{array}
       \right).
\end{eqnarray}
This block Hamiltonian defines a Weyl
semimetal with two Weyl nodes located at $\mathbf{k}^{\pm}_0$. (The lower block
$H^{-}_{2\times 2}$ has a similar form, except that $k_x$ is replaced by $-k_x$.)
It is well known \cite{Okugawa:2014ina,Vishwanath} that such a Weyl semimetal has the
surface Fermi arc connecting the Weyl nodes of opposite
chirality at $\mathbf{k}^{+}_0$ and $\mathbf{k}^{-}_0$. Because of the sign difference,
$k_x \to -k_x$, the chiralities of the states near the Weyl nodes at $\mathbf{k}^{\pm}_0$
are opposite for the upper and lower block Hamiltonians. Thus, the complete
$4\times 4$ block diagonal Hamiltonian $H_{4\times 4}(\alpha=0)$ describes two
superimposed copies of Weyl semimetal with two pairs of overlapping nodes. Since
the opposite chirality Weyl nodes coincide exactly in momentum space, they
effectively give rise to a pair of Dirac points at $\mathbf{k}^{\pm}_0$. At the same
time, because the opposite chirality nodes come from two different Weyl copies, they cannot
annihilate and cannot form topologically trivial Dirac points. In fact, the corresponding
approximate model describes a $\mathbb{Z}_2$ Weyl semimetal \cite{Gorbar:2014sja}.
The nontrivial topological properties, associated with the underlying $\mathbb{Z}_2$
Weyl semimetal structure, ensure that the resulting Dirac semimetal possesses
surface Fermi arcs.

It is easy to show that  the existence of the $\mathbb{Z}_2$ Weyl semimetal structure in the absence of
$B(\mathbf{k})$ is connected with the continuous symmetry $\mathrm{U}_{+}(1)\times \mathrm{U}_{-}(1)$
of the approximate Hamiltonian $H_{4\times 4}(\alpha=0)$. This symmetry describes
independent phase transformations of the spinors that correspond to the up- and down-block
Hamiltonians, $H^{+}_{2\times 2}$ and $H^{-}_{2\times 2}$, respectively.

For $B(\mathbf{k})\neq 0$, the continuous symmetry $\mathrm{U}_{+}(1)
\times \mathrm{U}_{-}(1)$ is broken down to its diagonal subgroup $\mathrm{U}_{\rm em}(1)$
that describes the usual charge conservation. However, the low-energy Hamiltonian
(\ref{low-energy-Hamiltonian}) with the momentum dependent mass function $B(\mathbf{k})
=\alpha k_zk^2_{+}$ possesses a ud-parity, defined by the following transformation
\cite{Gorbar:2014sja}:
\begin{equation}
U H_{-k_z} U^{-1}= H_{k_z} ,\quad \mbox{(ud-parity)},
\label{flavor-symmetry}
\end{equation}
where matrix $U$ has the following block diagonal form: $U\equiv \mbox{diag}(I_2,-I_2)$
and $I_2$ is the $2\times 2$ unit matrix. For the Hamiltonian to be symmetric
under the ud-parity, it is crucial that the mass function $B(\mathbf{k})$ changes its sign when
$k_z\to -k_z$ [while the functions $\epsilon_0(\mathbf{k})$ and $M(\mathbf{k})$ in the diagonal
elements do not change their signs]. In the special case of a momentum independent mass
function, such a discrete symmetry does not exist.

As was argued in Ref.~\cite{Gorbar:2014sja}, the existence of the noncommuting time-reversal
and ud-parity symmetries implies that the $\mathrm{A_3 Bi}$ semimetal is, in fact, a
$\mathbb{Z}_2$ Weyl semimetal. In such a semimetal, all quasiparticle states can be
split into two separate groups, labeled by the eigenvalues $\chi=\pm 1$ of $U_{\chi}
=U\Pi_{k_z}$, where $\Pi_{k_z}$ is the operator that changes the sign of the $z$
component of momentum, $k_z \to -k_z$. Effectively, each group of states defines
a Weyl semimetal with a broken time-reversal symmetry. The corresponding symmetry
is preserved in the complete theory, in which the two copies of Weyl semimetals are
superimposed.

The  $\mathbb{Z}_2$ Weyl semimetal structure of $\mathrm{A_3 Bi}$
($\mathrm{A}=\mathrm{Na}, \mathrm{K}, \mathrm{Rb}$) is also supported
by the explicit calculation of the  Berry connection and the Berry curvature in
each Weyl sector described \cite{Gorbar:2014sja}. In particular, the corresponding
results for the curvature in the momentum space reveal a clear dipole structure.
It is natural, that each Weyl sector, described by quasiparticle states with a fixed
eigenvalue of $U_\chi$, should give rise to Fermi arcs connecting the pairs of Weyl
nodes at $\mathbf{k}^{\pm}_0$ . Moreover, such arcs should be topologically
protected and could not be removed by small perturbations of model parameters.

In our discussion of Fermi arcs below, it will be also useful to take into account that
there exists yet another discrete symmetry defined by the following transformation:
\begin{eqnarray}
\tilde{U} H_{-k_x} \tilde{U}^{-1} &=& H_{k_x},
\label{second-discrete-symmetry}
\end{eqnarray}
where
\begin{equation}
\tilde{U} = \left(
       \begin{array}{cccc}
           0 & 0 & 1 & 0 \\
           0 & 0 & 0 & 1  \\
           1 & 0 & 0 & 0 \\
           0 & 1 & 0 & 0
         \end{array}
       \right).
\label{U-prime-matrix}
\end{equation}
It is interesting to note that the product of the $U_{\chi}$ and $\tilde{U}\Pi_{k_x}$ transformations
$U_{\chi}\tilde{U}\Pi_{k_x}=T\Pi_{k_x}\Pi_{k_z}$ is also a symmetry of the low-energy
Hamiltonian (\ref{low-energy-Hamiltonian}). The symmetry $T\Pi_{k_x}\Pi_{k_z}$ is
related to the time-reversal symmetry. This follows from the fact that $K\Pi_{k_y}$
is also the symmetry of the low-energy Hamiltonian (\ref{low-energy-Hamiltonian}).
Together the operators $U_{\chi}$, $\tilde{U}\Pi_{k_x}$, and $T\Pi_{k_x}\Pi_{k_z}$
form a non-commutative discrete group.

Hamiltonian (\ref{low-energy-Hamiltonian}) is rather complicated, therefore, the corresponding
analytic calculations of its surface Fermi states are quite involved and not much revealing.
Therefore, our general strategy in analyzing these states will be to start from a simplified
model and then move forward to the realistic model by adding step-by-step the necessary
missing pieces.

\section{Surface Fermi arcs in simplified $2\times 2$ model}
\label{sec:2x2Model}

In order to get an insight into the structure of the surface Fermi arcs in the low-energy
model described by Hamiltonian (\ref{low-energy-Hamiltonian}), it is instructive to
first study the surface Fermi arcs in a simplified $2\times 2$ model, given by one of the diagonal
blocks, e.g., $H^{+}_{2\times 2}$ in Eq.~(\ref{Hamiltonians_+_New}). (The solutions for the other
block Hamiltonian, $H^{-}_{2\times 2}$, can be obtained simply by changing $k_x \to -k_x$.)
For completeness, we will also include the term $\epsilon_0(\mathbf{k})$ proportional to the
unit matrix, which is present in the low-energy Hamiltonian. Thus, our model $2\times 2$
Hamiltonian reads
\begin{equation}
H_{2\times 2} = \epsilon_0(\mathbf{k})+H^{+}_{2\times 2}=\epsilon_0(\mathbf{k})
 +\left(
         \begin{array}{cc}
           M_0-M_1k^2_z-M_2(k^2_x+k^2_y) & A(k_x+ik_y) \\
           A(k_x-ik_y) & -\left[M_0-M_1k^2_z-M_2(k^2_x+k^2_y)\right]  \\
         \end{array}
       \right).
\label{block-model}
\end{equation}
Before proceeding to the analysis, it is convenient to perform a unitary transformation,
$\tilde{H}_{2\times 2} \equiv U_{y}^{-1}H_{2\times 2}U_{y}$, where $U_{y} = \frac{1}{\sqrt{2}}
\left(I_2+i\sigma_y\right)$. The transformed Hamiltonian has the following explicit form:
\begin{equation}
\tilde{H}_{2\times 2} = \epsilon_0(\mathbf{k})+ \left[\gamma\left(k_z^2 - m\right)-M_2(k^2_x+k_y^2)\right]\sigma_x
-vk_x\sigma_z-vk_y\sigma_y, 
\label{model-xxx}
\end{equation}
where we introduced the notations similar to those in Ref.~\cite{Okugawa:2014ina}: $v=A$ and
$\gamma=-M_1$.

To study the surface Fermi arcs, we will assume that the surface of a semimetal is
at $y=0$. The semimetal itself is in the upper $y>0$ (lower $y<0$) half-space when
we describe the surface arc states on the bottom (top) surface. (Of course, in the absence of
any effects that break the inversion symmetry $k_y\to -k_y$ explicitly, the two cases
will be related by a simple symmetry transformation.) Without loss of generality, we
will concentrate primarily on the bottom surface states. The boundary condition on
the semimetal surface will be imposed by replacing the parameter $m$ with the
$-\tilde{m}$ on the vacuum side of the boundary and taking the limit $\tilde{m}\to
\infty$ \cite{Okugawa:2014ina}. From a physics viewpoint, such a replacement is the simplest
way to prevent quasiparticle from escaping into the vacuum.

Taking into account that the Fermi arc states should be localized at the $y=0$
boundary, let us rewrite Hamiltonian (\ref{model-xxx}) in the following form:
\begin{equation}
\tilde{H}_{2\times 2} =  \left(
     \begin{array}{cc}
       C_0+C_1k^2_z+C_2(k^2_x-\partial_y^2)-vk_x & \gamma\left(k_z^2 - m\right)-M_2(k^2_x-\partial_y^2)+v\partial_y \\
       \gamma\left(k_z^2 - m\right)-M_2(k^2_x-\partial_y^2)-v\partial_y & C_0+C_1k^2_z+C_2(k^2_x-\partial_y^2)+vk_x \\
     \end{array}
   \right),
\label{block-model-U}
\end{equation}
where, for the convenience of further derivations, we replaced $k_y\equiv -i\partial_y$.

\subsection{Simplified model with $C_2= M_2=0$}
\label{sec:2x2ModelC2=0M2=0}

\begin{figure}[!ht]
\begin{center}
\includegraphics[width=0.48\textwidth]{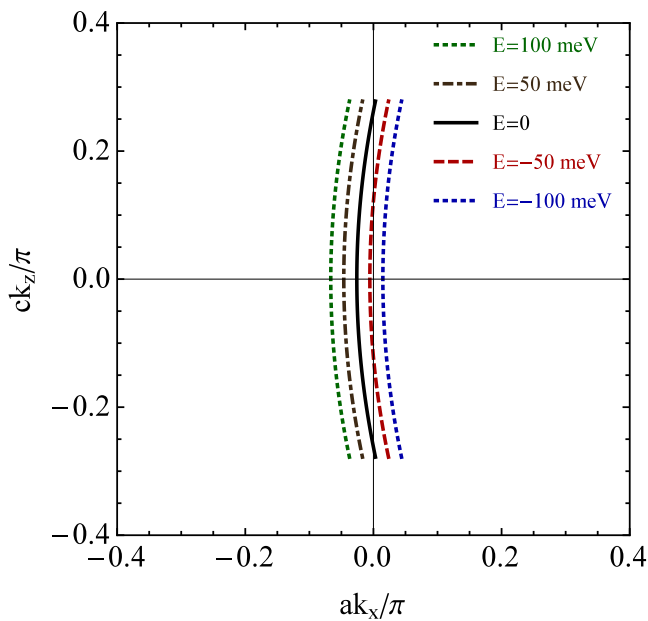}
\caption{(Color online) The bottom surface Fermi arcs for several different values of the
Fermi energy in a simplified two-component model, described by Hamiltonian (\ref{block-model-U})
with $C_2= M_2=0$. The analytical form of the arcs is given in Eq.~(\ref{kx-equation-1}).}
\label{fig:Fermi-arc-00}
\end{center}
\end{figure}

We will see in what follows that the presence of the terms with the second derivative
with respect to $y$ in Hamiltonian (\ref{block-model-U}) leads to many technical complications
and makes the analysis rather involved. Therefore, to set up the stage, in this subsection we
start our analysis in an even more simplified model, described by
Hamiltonian (\ref{block-model-U}) with $C_2$ and $M_2$ set to zero.
Then, by introducing the two-component spinor $\Psi = \left(\psi_1 , \psi_2 \right)^{T}$,
we see that the eigenvalue problem $(\tilde{H}_{2\times 2}  -E)\Psi = 0$ is equivalent
to the following system of equations:
\begin{eqnarray}
\label{psi-equation-1-00}
\left(-v k_x+C_1k_z^2+C_0\right)\psi_1  + \left[v \partial_y +\gamma k_z^2-\gamma m(y)\right]\psi_2 &=& E \psi_1 , \\
\label{psi-equation-2-00}
\left(v k_x+C_1k_z^2+C_0\right)\psi_2 + \left[-v \partial_y +\gamma k_z^2-\gamma m(y)\right]\psi_1 &=& E \psi_2 .
\end{eqnarray}
Here $m(y)=m\theta(y)-\tilde{m}\theta(-y)$, where $\theta(y)$ is the step function.
Recall that, by assumption, the boundary condition
at $y=0$ is enforced by taking the limit $\tilde{m}\to \infty$ on the vacuum side ($y<0$). Formally,
Eqs.~(\ref{psi-equation-1-00}) and (\ref{psi-equation-2-00}) have the following surface state solutions:
\begin{equation}
\Psi_{1}(y)=\left(
\begin{array}{c}
N_{1} e^{\frac{\gamma}{v} \int^y dy^{\prime} \left[k_z^2- m(y^{\prime})\right]} \\
0
\end{array}
\right),
\qquad
\Psi_{ 2}(y)=\left(
\begin{array}{c}
0 \\
N_{2} e^{-\frac{\gamma}{v} \int^y dy^{\prime} \left[k_z^2- m(y^{\prime})\right]} \\
\end{array}
\right),
\label{psi_1-00}
\end{equation}
In the region occupied by the semimetal ($y>0$), the solution $\Psi_1(y)$ is normalizable only for
$k_z^2- m<0$, while the solution $\Psi_2(y)$ is normalizable only for $k_z^2 - m>0$. However, on the vacuum
side ($y<0$), only $\Psi_1(y)$ is normalizable. The dispersion relation for this normalizable
surface state solution follows from Eq.~(\ref{psi-equation-1-00}). It is given by
\begin{equation}
E = - v k_x+C_1 k_z^2 + C_0.
\label{E-00}
\end{equation}
By making use of this relation, we derive the equation for the bottom surface Fermi arc
in the transverse $k_xk_z$ plane,
\begin{equation}
k_x=-\frac{E-C_1 k_z^2 - C_0}{v}.
\label{kx-equation-1}
\end{equation}
It is instructive to compare this surface Fermi arc with that in the model of Ref.~\cite{Okugawa:2014ina},
where $C_1=0$. While the surface Fermi arcs run between $k_z=-\sqrt{m}$ and $k_z=\sqrt{m}$ in both
models, the arcs in the model of Ref.~\cite{Okugawa:2014ina} do not depend on the momentum
$k_z$. This is in contrast to the surface Fermi arc in Eq.~(\ref{kx-equation-1}), for which $k_x$ is a quadratic
function of $k_z$. Thus, we see that the presence of the quadratic in $k_z$ term in the diagonal
component of Hamiltonian (\ref{block-model-U}) produces a nonzero {\it curvature} of the surface
Fermi arcs in momentum space. For illustration, several surface Fermi arcs for different values
of the Fermi energy are shown in Fig.~\ref{fig:Fermi-arc-00}. The arcs have parabolic shapes.
The corresponding arcs in the model of Ref.~\cite{Okugawa:2014ina} would be given by straight lines.

%%%%%%%%%%%%%%%%%%
%%%%%%%%%%%%%%%%%%

Before concluding this section, let us note that the solution $\Psi_2(y)$ in Eq.~(\ref{psi_1-00})
describes Fermi arcs on the top surface. We find from Eq.~(\ref{psi-equation-2-00}) that the
corresponding dispersion relation is given by $E = v k_x+C_1 k_z^2 + C_0$. Let us also
note in passing that there exists another set of the (top and bottom) Fermi arcs for the lower
block Hamiltonian, $H^{-}_{2\times 2}$. The corresponding arcs are obtained from the solutions
for the upper block Hamiltonian, $H^{+}_{2\times 2}$, by making the replacement
$k_x\rightarrow-k_x$.

\subsection{The case with $C_2\neq 0$ and $M_2 \neq 0$}
\label{sec:2x2ModelC2M2}

Let us now consider the general case with $C_2\neq 0$ and $M_2 \neq 0$. By noting
that the Hamiltonian in Eq.~(\ref{block-model-U}) contains second derivatives with respect
to $y$, the eigenvalue problem $(\tilde{H}_{2\times 2} -E)\Psi = 0$ becomes more complicated.
In the semimetal ($y>0$), it is equivalent to the following system of coupled equations:
\begin{eqnarray}
\label{psi-equation-1-01}
\left[ C_2(k_x^2 - \partial_y^2)-v k_x+C_1k_z^2+C_0\right]\psi_1
+ \left[-M_2(k_x^2 - \partial_y^2)+v \partial_y +\gamma k_z^2
-\gamma m\right]\psi_2 &=& E \psi_1 , \\
\label{psi-equation-2-01}
\left[ C_2(k_x^2 - \partial_y^2)+v k_x+C_1k_z^2+C_0\right]\psi_2
+ \left[-M_2(k_x^2 - \partial_y^2)-v \partial_y +\gamma k_z^2
-\gamma m\right]\psi_1 &=& E \psi_2 .
\end{eqnarray}
On the vacuum side ($y<0$), the corresponding set of equations has the same form, but with
$m$ replaced by $-\tilde{m}$. At the vacuum-semimetal interface ($y=0$), the wave functions
and their derivatives should satisfy the conditions of continuity, see Eqs.~(\ref{cross-liking0})
through (\ref{cross-liking3}) in Appendix~\ref{AppExtra1}.

\begin{figure*}
\begin{center}
\includegraphics[width=0.48\textwidth]{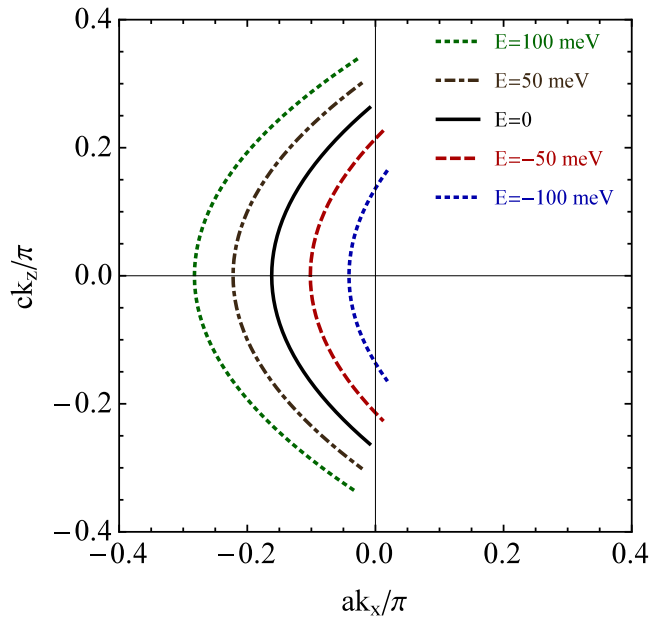}
\caption{(Color online) The bottom surface Fermi arcs (\ref{Q1=Q2_alpha=0}) for several values
of the Fermi energy in a two-component model, described by Hamiltonian (\ref{block-model-U})
with $C_2\neq 0 $ and $M_2\neq 0$.}
\label{fig:Fermi-Arc-5Es}
\end{center}
\end{figure*}

The key details of the derivation of the surface Fermi arc solutions are presented in
Appendix~\ref{AppExtra1}. On the semimetal side, the spinor structure of the solution
takes the following form:
\begin{equation}
\Psi_{y>0}(y)
=\sum_{i=1}^{2}\left(\begin{array}{r}
a_i \\
b_i
\end{array}\right)e^{- p_i y} ,
\end{equation}
where the explicit expressions for the exponents are given in Eq.~(\ref{p1p2-two-component}).
Note that the exponents take real values in the case of surface Fermi arc states.
The condition of existence of nontrivial surface Fermi arc solutions is given by
\begin{equation}
\frac{-C_2(p_1^2-k_x^2)+C_1k_z^2+C_0-E-v k_x}{-M_2(p_1^2-k_x^2)-\gamma (k_z^2-m)+vp_1}
=\frac{-C_2(p_2^2-k_x^2)+C_1k_z^2+C_0-E-v k_x}{-M_2(p_2^2-k_x^2)-\gamma (k_z^2-m)+vp_2}.
\label{Q1=Q2_alpha=0}
\end{equation}
This equation defines the functional dependence $k_z(k_x)$ for the possible surface
Fermi arc states. A numerical study shows that nontrivial solutions exist only in a finite range of
energies, i.e., $-0.168~\mbox{eV} \lesssim E \lesssim 0.373~\mbox{eV}$. Several solutions
for different values of the energy are shown in Fig.~\ref{fig:Fermi-Arc-5Es}.
The results of the numerical analysis show that the following condition is satisfied:
$b_1/a_1= b_2/a_2 = 0.5115$ for all solutions. It is worth noting that the $E=0$ surface
Fermi arc in Fig.~\ref{fig:Fermi-Arc-5Es} appears to be almost identical to the
corresponding arc, obtained by a very different method in Ref.~\cite{Fang},
see Fig.~3c in that paper.

So far, we considered the arc states only for one of the two-component block Hamiltonians,
defined in Eq.~(\ref{block-model}). Similar solutions also exist for the lower two-component
block Hamiltonian, $\epsilon_0(\mathbf{k})+H^{-}_{2\times 2}$. It is straightforward to show
that the solutions to the eigenvalue problem for the lower block are the same as for the upper
one, after one makes the replacement $k_x\rightarrow-k_x$. Graphically, these solutions are
mirror images of the arcs in Fig.~\ref{fig:Fermi-Arc-5Es}.

Before concluding this subsection, let us also note that the description of the Fermi arc states
on the top surface is similar to the bottom ones. By assuming that Weyl semimetal is at $y<0$
and the vacuum is at $y>0$, the appropriate boundary conditions are implemented by using the
$y$-dependent parameter $m(y)=m\theta(-y)-\tilde{m}\theta(y)$ and taking the limit $\tilde{m}
\to \infty$ at the end. Up to a reflection $k_x\to -k_x$, the corresponding final results for the
Fermi arcs on the top surface look similar to those on the bottom surface, shown in
Fig.~\ref{fig:Fermi-Arc-5Es}.

\subsection{Effective Hamiltonian for surface Fermi arc states}
\label{surface-Hamiltonian}

Following the usual approach in the studies of topological insulator \cite{Shen}, it may be natural 
to derive an effective Hamiltonian for the surface Fermi arc states.
The block Hamiltonians in the simplified model at hand can be naturally separated into two parts, i.e.,
$\tilde{H}^{\pm}_{2\times 2} = H_0+H^{\pm}_1$, where the zeroth order part $H_0$ corresponds to the
original Hamiltonian at $k_x=k_z=0$, i.e.,
\begin{equation}
H_0=\left(
     \begin{array}{cc}
       C_0-C_2\partial_y^2 & -\gamma m+M_2\partial_y^2+v\partial_y \\
       -\gamma m+M_2\partial_y^2-v\partial_y & C_0-C_2\partial_y^2 \\
     \end{array}
   \right).
\end{equation}
while $H_1$ contains all the terms with nontrivial dependence on $k_x$ and $k_z$, i.e.,
\begin{equation}
H^{\pm}_1=\left(
     \begin{array}{cc}
       C_1k^2_z+C_2k^2_x \mp v k_x & \gamma k^2_z -M_2k^2_x \\
       \gamma k^2_z -M_2k^2_x & C_1k^2_z +C_2k^2_x \pm v k_x \\
     \end{array}
   \right).
\label{block-model-M2}
\end{equation}
As in the previous analysis, we used $k_y\equiv -i\partial_y$. To start with, we have to
solve the eigenvalue problem with the zeroth order Hamiltonian, $H_0\Psi_0=\lambda\Psi_0$.
By following the same approach as in Appendix~\ref{AppExtra1}, but with $k_x=k_z=0$, we find
straightforwardly the explicit solutions for the surface Fermi arcs $\Psi_0$. The corresponding
energy parameter is found to be $\lambda = -0.13425~\mbox{eV}$. Then, the effective Hamiltonian
for the surface states is obtained by integrating over the perpendicular direction $y$, i.e.,
\begin{eqnarray}
H^{\pm}_{\rm surf} &=&  \lambda+\int_0^{\infty} dy \Psi^{\dag}_0H_1\Psi_0
=\lambda+ C_1k^2_z+C_2k_x^2 \mp v k_x\frac{1-Q^2}{1+Q^2}
+2(\gamma k^2_z -M_2k_x^2)\frac{Q}{1+Q^2}
\nonumber\\
&\approx & \lambda \mp v_{\rm surf} k_x +\gamma_{\rm surf} k^2_z.
\label{Hamiltonian-effective-1-M2}
\end{eqnarray}
where $Q\approx  0.5115$, $v_{\rm surf}\approx 1.440~\mbox{eV\,\AA}$ and $\gamma_{\rm surf}
\approx 17.38~\mbox{eV\,\AA}^2$. Note that the quadratic term in $k_x$ vanishes after the model
parameters are used.

As is easy to check, the effective Hamiltonian in Eq.~(\ref{Hamiltonian-effective-1-M2}) reproduces
almost perfectly the shape of the Fermi arcs in the $k_xk_z$ plane. However, it does not contain
the information about the finite length of the arcs. We could explain this fact in part by pointing out
that the corresponding information is encoded in the terms quadratic in momenta $k_x$ and $k_z$.
When such terms are omitted from the zeroth order Hamiltonian $H_0$, the existence of the surface
states formally appears to be unconstrained. Therefore, the effective Hamiltonian in
Eq.~(\ref{Hamiltonian-effective-1-M2}) will be truly useful only when supplemented by its range of
validity in the $k_xk_z$ plane. This, however, seems to diminish its practical value because the
corresponding range depends on the energy.

\section{Fermi arcs in realistic model}
\label{sec:Realistic4x4Model}

In this section we will consider the complete low-energy theory described by Hamiltonian
(\ref{low-energy-Hamiltonian}) with $\alpha\neq0$. By performing a unitary transformation
in Eq.~(\ref{low-energy-Hamiltonian}), defined by $U_{y} = \frac{1}{\sqrt{2}}I_2
\otimes\left(I_2+i\sigma_y\right)$, we arrive at the following equivalent
form of the Hamiltonian:
\begin{eqnarray}
\tilde{H} &=& \left[C_2(k_x^2 - \partial_y^2)+C_1k_z^2+C_0\right] I_2\otimes I_2 -M_2(k_x^2 - \partial_y^2) I_2\otimes \sigma_x  \nonumber \\
&&+\left( \begin{array}{cccc}
      -v k_x  & v \partial_y +\gamma (k_z^2- m) & -\alpha
       k_z(k_x-\partial_y)^2 & 0 \\
       -v \partial_y +\gamma (k_z^2- m) &v k_x & 0 & \alpha
       k_z(k_x-\partial_y)^2 \\
       -\alpha k_z(k_x+\partial_y)^2 & 0 & v k_x  & v \partial_y +\gamma (k_z^2- m) \\
       0 & \alpha k_z(k_x+\partial_y)^2 & -v \partial_y +\gamma (k_z^2- m) & -v k_x \\
       \end{array}
       \right),
       \label{Wang-Hamiltonian-2}
\end{eqnarray}
By introducing the spinor wave function $\Psi = \left(\psi_1 , \psi_2 , \psi_3 , \psi_4 \right)^{T}$,
we reduce the eigenvalue problem $(\tilde{H}  -E)\Psi = 0$ in the semimetal ($y>0$)
to the following system of equations:
\begin{eqnarray}
\label{psi-equation-1-four}
\left[ C_2(k_x^2 - \partial_y^2)-v k_x+C_1k_z^2+C_0-E\right]\psi_1
+\left[-M_2(k_x^2 - \partial_y^2)+v \partial_y +\gamma k_z^2 -\gamma m\right]\psi_2
- \alpha k_z (k_x-\partial_y)^2\psi_3  =0, &&  \\
\label{psi-equation-2-four}
\left[-M_2(k_x^2 - \partial_y^2)-v \partial_y +\gamma k_z^2-\gamma m\right]\psi_1
+ \left[ C_2(k_x^2 - \partial_y^2)+v k_x+C_1k_z^2+C_0 -E\right]\psi_2
+ \alpha k_z (k_x-\partial_y)^2\psi_4  =0, &&  \\
\label{psi-equation-3-four}
-\alpha k_z (k_x+\partial_y)^2\psi_1
+\left[ C_2(k_x^2 - \partial_y^2)+v k_x+C_1k_z^2+C_0-E\right]\psi_3
+\left[-M_2(k_x^2- \partial_y^2)+v \partial_y +\gamma k_z^2-\gamma m\right]\psi_4 =0, &&  \\
\label{psi-equation-4-four}
\alpha k_z (k_x+\partial_y)^2\psi_2
+\left[-M_2(k_x^2 - \partial_y^2)-v \partial_y +\gamma k_z^2-\gamma m\right]\psi_3
+\left[ C_2(k_x^2 - \partial_y^2)-v k_x+C_1k_z^2+C_0-E\right]\psi_4  =0 . &&
\end{eqnarray}
On the vacuum side ($y<0$), the corresponding set of equations has the same form, but with
$m$ replaced by $-\tilde{m}$. The corresponding full set of equations should be also supplemented
by the conditions of continuity of the wave functions and their derivatives across the vacuum-semimetal
interface at $y=0$, see Eqs.~(\ref{cross-liking-four1}) through (\ref{cross-liking-four5}) in
Appendix~\ref{AppExtra2}.

As shown in Appendix~\ref{AppExtra2}, the spinor structure of the solution on the semimetal side
takes the form:
\begin{equation}
\Psi_{y>0}(y)
=\sum_{i=1}^{2}\left(\begin{array}{r}
a_i \\
b_i  \\
c_i  \\
d_i
\end{array}\right)e^{- p_i y} ,
\end{equation}
where the explicit expressions for the exponents are given in Eq.~(\ref{p12-four-comp}).
In the case of surface Fermi arc solutions, the exponents take real values.
A nontrivial solution exists when the following condition is satisfied:
\begin{equation}
\left(Q_1^{+}  -Q_2^{+}\right)\left(Q_1^{-}  -Q_2^{-}\right)
-\left(T_1^{+}  -T_2^{+}\right)\left(T_1^{-}  -T_2^{-}\right)=0,
\label{key-equation}
\end{equation}
where, by definition, $Q_{i}^{\pm} \equiv Q(p_i,\pm k_x)$ and $T_{i}^{\pm} \equiv T(p_i,\pm k_x)$,
and the functions $Q(p,k_x)$ and $T(p,k_x)$  are defined in Eqs.~(\ref{def-Qpkx}) and (\ref{def-Tpkx}),
respectively.

By taking into account that $T(p,k_x)$ vanishes at $\alpha = 0$, one finds that the above condition
reduces to its analog in Eq.~(\ref{Q1=Q2_alpha=0}) in the two-component model. Indeed, a nontrivial
solution exists in the model with the two-component upper (lower) block Hamiltonian when $Q_1^{+} =Q_2^{+}$
($Q_1^{-} =Q_2^{-}$) is satisfied. We would like to emphasize that the classification of the arc states
remains essentially the same also in a general case with $\alpha \neq 0$. However, because of the mixing
between the upper and lower block Hamiltonians, the arcs are labeled by the eigenvalues of the $U_{\chi}$
operator, see Appendix~\ref{AppA}. The eigenstates with $\chi=+1$ ($\chi=-1$) are the generalizations
of the arcs from the upper (lower) block Hamiltonian.

\begin{figure*}[!ht]
\begin{center}
\includegraphics[width=0.47\textwidth]{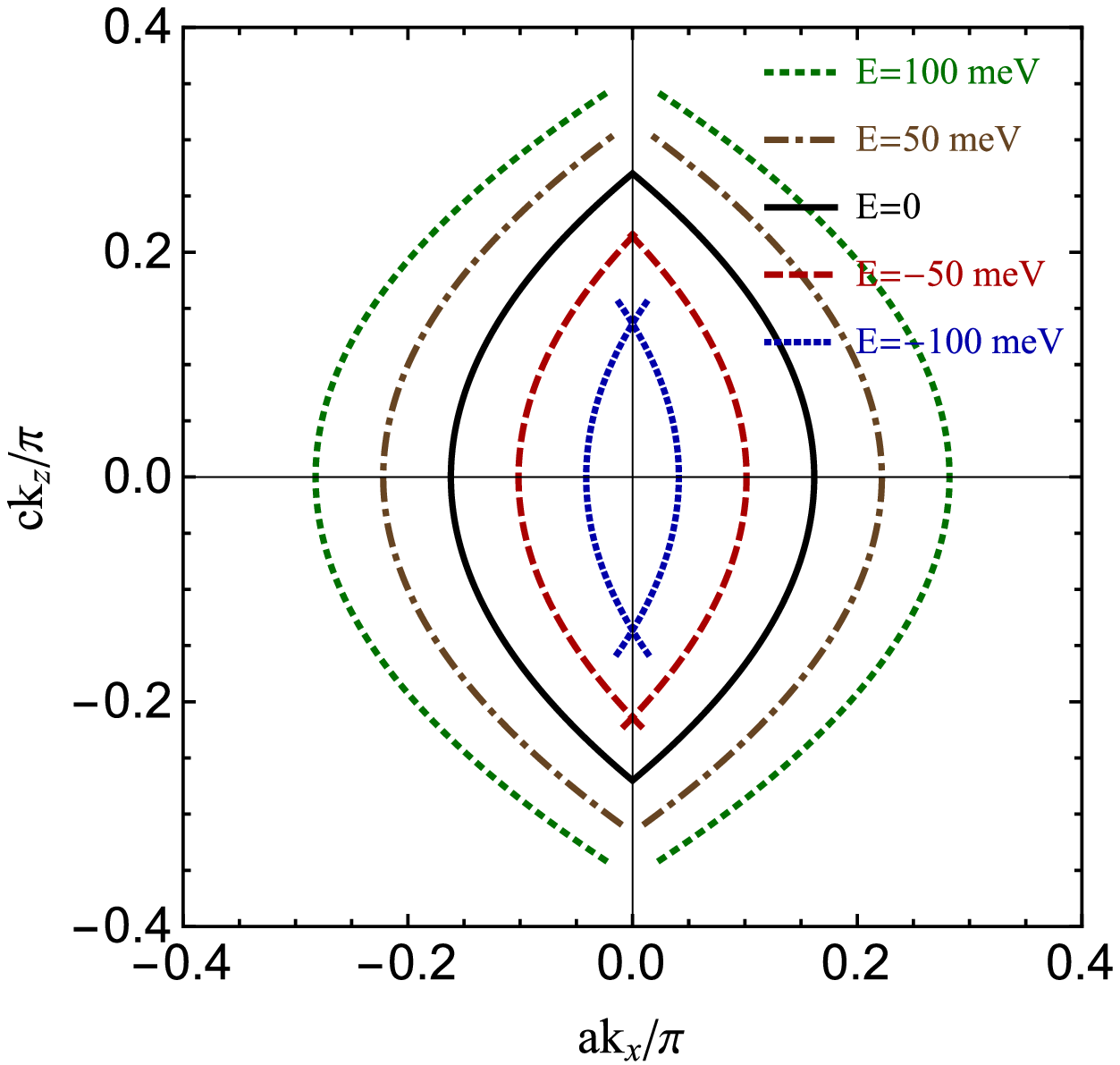}
\includegraphics[width=0.47\textwidth]{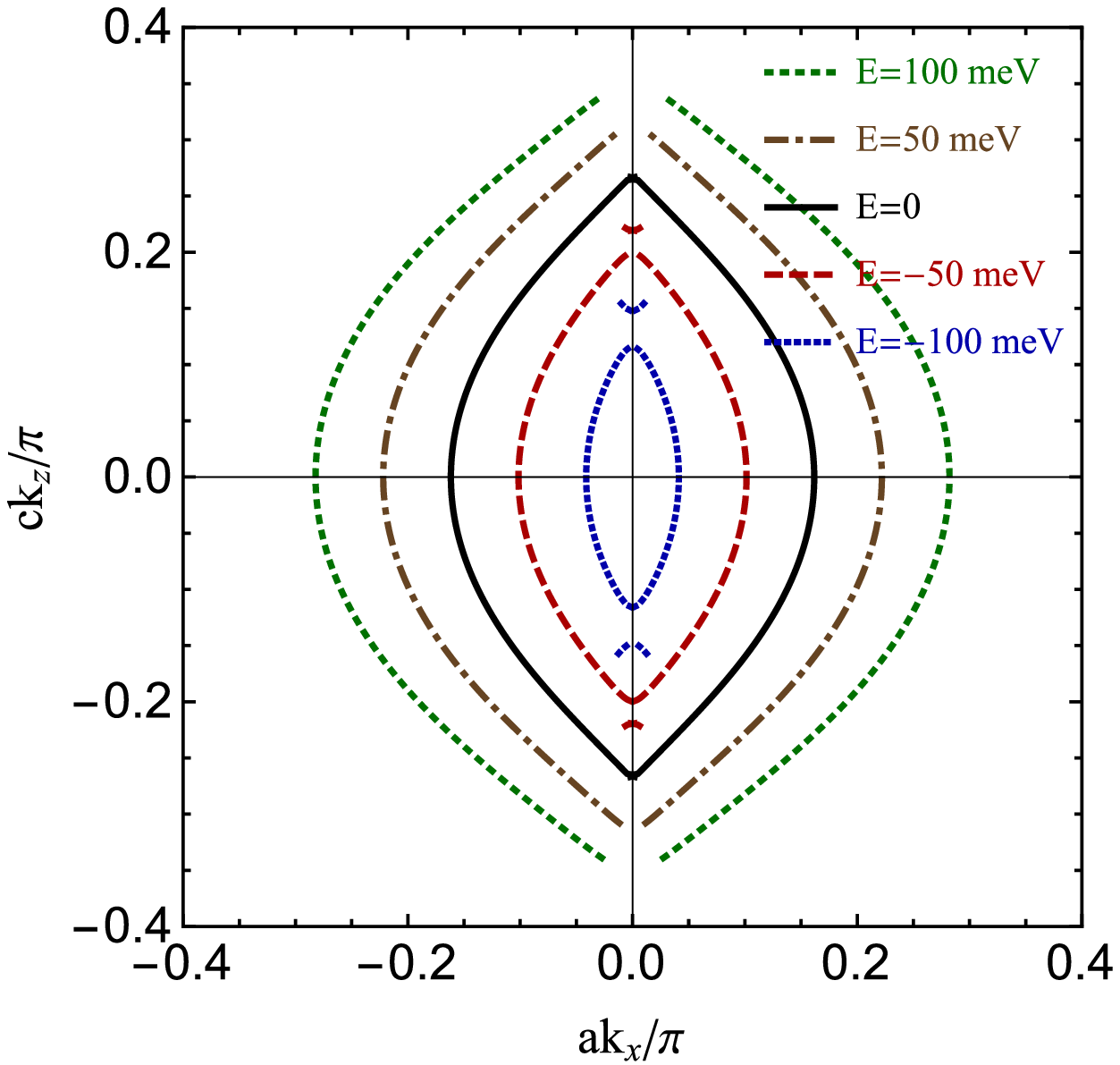}
\caption{(Color online) The Fermi arcs solutions in the plane of transverse momenta
for $\alpha=1~\mbox{eV\,\AA}^3$ (left panel) and $\alpha=50~\mbox{eV\,\AA}^3$ (right panel).}
\label{fig:Fermi-Arc-5Es_alpha}
\end{center}
\end{figure*}

The numerical results for the surface Fermi arc states are shown in
Fig.~\ref{fig:Fermi-Arc-5Es_alpha} for $\alpha=1~\mbox{eV\,\AA}^3$ (left panel) and
$\alpha=50~\mbox{eV\,\AA}^3$ (right panel). At fixed energy, there are two
surface Fermi arcs related to two different sectors of the $\mathrm{A_3Bi}$ (A=Na, K, Rb)
compounds with definite eigenvalue of $U_{\chi}$. One can check that the wave functions
that describe these surface Fermi arcs are related to each other by means of the
$\tilde{U}\Pi_{k_x}$ transformation, see Appendix~\ref{AppA}. By comparing these results
with those in the two-component model, see Fig.~\ref{fig:Fermi-Arc-5Es}, we find that the quantitative
effect of a nonzero $\alpha$ on the Fermi arcs is small even when $\alpha$ is moderately
large. The only qualitative effect due to $\alpha$ is a reconnection of the pair of arcs
(from predominantly up and predominantly down sectors) at negative values of the
Fermi energy. The underlying physics of such an effect is likely to be connected with
the loss of the chirality as a good quantum number for quasiparticles away from the
Dirac/Weyl nodes. Because of the discrete ud-parity, which is preserved even at
large values of $\alpha$, there are still two sectors of the theory and there are
still small nontrivial arcs present, as we see from the right panel of Fig.~\ref{fig:Fermi-Arc-5Es_alpha}.
It will be interesting to explore whether the reconnection of the pairs of arcs would
also appear in the microscopic theory. It may well be an artifact of the low-energy
theory used here.

\section{Fermi arcs and weak breaking of time-reversal symmetry}
\label{sec:DynamicalParametersModel}

As we discussed in detail in Sec.~\ref{sec:symmetries}, the low-energy effective Hamiltonian
(\ref{low-energy-Hamiltonian}) is invariant under the time-reversal and inversion symmetries.
Moreover, these symmetries play an important role in defining the physical properties of
$\mathrm{A_3Bi}$ semimetals. Thus, it is natural to ask about possible effects on the
structure (and perhaps even the existence) of surface Fermi arcs due to breaking of these
symmetries. From the physics viewpoint, for example, the corresponding discrete symmetries
could be broken explicitly by magnetic doping or an external magnetic field.

In order to study the symmetry breaking effects, we will add to the low-energy Hamiltonian
(\ref{low-energy-Hamiltonian}) two additional terms controlled by parameters $m_1$ and
$\tilde{\mu}_1$:
\begin{eqnarray}
H_{\rm sb} &=& H(\mathbf{k}) - \left(
                              \begin{array}{cc}
                                \tilde{\mu}_1I_{2}+\sigma_z \gamma m_1 & 0 \\
                                0 & -\tilde{\mu}_1I_{2}-\sigma_z \gamma m_1 \\
                              \end{array}
                            \right).
       \label{Wang-Hamiltonian-0-3}
\end{eqnarray}
By analyzing the Schwinger--Dyson equation for the quasiparticle propagator in $\mathrm{A_3Bi}$ semimetals in a magnetic field,
we found that these terms are indeed perturbatively generated. Alternatively, these terms can be induced by magnetic doping.
The value of $\tilde{\mu}_1$ could be interpreted as a mismatch between the chemical
potentials of quasiparticle states in the Weyl sectors of the theory. The value of $m_1$
is a mismatch of the parameter $m$ that determines the chiral shift in the two sectors.
This means that whenever these symmetry breaking parameters appear, the $\mathbb{Z}_2$
Weyl semimetal will get automatically transformed into a true Weyl semimetal with four
non-degenerate Weyl nodes.

By performing a unitary transformation in Eq.~(\ref{Wang-Hamiltonian-0-3}), defined by matrix
$U_{y} = \frac{1}{\sqrt{2}}I_{2}\otimes\left(I_{2}+i\sigma_y\right)$, we arrive at the following
equivalent Hamiltonian:
\begin{eqnarray}
&&\tilde{H}_{\rm sb} =\left[C_2(k_x^2 - \partial_y^2)+C_1k_z^2+C_0\right] I_2\otimes I_2 -M_2(k_x^2 - \partial_y^2) I_2\otimes \sigma_x  \nonumber \\
&&+\left( \begin{array}{cccc}
       -v k_x -\tilde{\mu}_1  & v \partial_y +\gamma \left(k_z^2- m-m_1\right) & -\alpha k_z(k_x-\partial_y)^2 & 0 \\
        -v \partial_y +\gamma\left( k_z^2- m-m_1\right) & v k_x -\tilde{\mu}_1 & 0 & \alpha k_z(k_x-\partial_y)^2 \\
       -\alpha k_z(k_x+\partial_y)^2 & 0 &  v k_x+\tilde{\mu}_1  & v \partial_y +\gamma\left( k_z^2-
       m+m_1\right) \\
       0 & \alpha k_z(k_x+\partial_y)^2 & -v \partial_y +\gamma\left( k_z^2- m+m_1\right) &  -v
       k_x+\tilde{\mu}_1 \\
       \end{array}
       \right). \nonumber \\
       \label{Wang-Hamiltonian-3}
\end{eqnarray}
It is straightforward, although tedious to repeat the same analysis as in Sec.~\ref{sec:Realistic4x4Model}.

The general surface state solution is of the same type, i.e., $\Psi_{y>0} (y) = \Psi_0 e^{- p y}$ ,
where $\Psi_0\equiv (a,b,c,d)^T$ is a constant spinor. However, the characteristic equation is
considerably more complicated,
\begin{eqnarray}
&&\left\{\left[ -C_2(p^2-k_x^2)+C_1k_z^2+C_0-\tilde{\mu}_1-E\right]^2
-\left[M_2(p^2-k_x^2) +\gamma (k_z^2-m-m_1)\right]^2+v^2 (p^2-k_x^2)- \alpha^2 k_z^2(p^2-k_x^2)^2\right\} \nonumber\\
&&\times\left\{\left[ -C_2(p^2-k_x^2)+C_1k_z^2+C_0+\tilde{\mu}_1-E\right]^2
-\left[M_2(p^2-k_x^2) +\gamma (k_z^2-m+m_1)\right]^2+v^2 (p^2-k_x^2)- \alpha^2 k_z^2(p^2-k_x^2)^2\right\}\nonumber\\
&&+4 \alpha^2 k_z^2(p^2-k_x^2)^2\left(\tilde{\mu}_1^2- \gamma^2 m_1^2\right)=0 .
\label{char-eq-2}
\end{eqnarray}
The important effect of the symmetry breaking terms with nonzero $m_1$ and $\tilde{\mu}_1$ is
that the new characteristic equation has {\em four} (instead of two degenerate) pairs of distinct
solutions: $p=\pm p_i$, with $i=1,2,3,4$. The general spinor solution in the semimetal takes
the following form:
\begin{equation}
\Psi_{y>0}(y)
=\sum_{i=1}^{4}\left(\begin{array}{r}
a_i \\
b_i  \\
c_i  \\
d_i
\end{array}\right)e^{- p_i y} .
\end{equation}
By making use of the equation of motion, the components $b_i$ and $d_i$ can be expressed
in terms of $a_i$ and $c_i$,
\begin{eqnarray}
b_i &=&  \frac{ -C_2(p_i^2-k_x^2)+C_1k_z^2+C_0-\tilde{\mu}_1-E-v k_x}
{-M_2(p_i^2-k_x^2)-\gamma\left( k_z^2- m-m_1\right) +v p_i } a_i
- \frac{ \alpha k_z (p_i+k_x)^2}
{-M_2(p_i^2-k_x^2)-\gamma\left( k_z^2- m-m_1\right) +v p_i }  c_i  ,\\
d_i &=&   - \frac{ \alpha k_z (p_i-k_x)^2}
{-M_2(p_i^2-k_x^2)-\gamma\left( k_z^2- m+m_1\right) +v p_i } a_i
+\frac{ -C_2(p_i^2-k_x^2)+C_1k_z^2+C_0+\tilde{\mu}_1-E-v k_x}
{-M_2(p_i^2-k_x^2)-\gamma\left( k_z^2- m+m_1\right) +v p_i }  c_i .
\end{eqnarray}
In order to avoid a possible confusion, let us emphasize that the remaining two components
$a_i$ and $c_i$ are not independent, but fixed unambiguously for each $p_i$. The final
solutions for the Fermi arcs are determined after all four independent parameters (e.g.,
$a_i$ with $i=1,2,3,4$) are fixed by satisfying the continuity conditions for the wave function
at the surface of the semimetal. The corresponding solutions can be obtained by numerical
methods.

To slightly simplify the analysis, let us consider a special case of vanishing $\alpha$ in more
detail. In this case, the states from the two-component upper and lower block Hamiltonians
decouple. Also, the characteristic equation factorizes, effectively giving two separate
equations, i.e.,
\begin{eqnarray}
\left[ -C_2(p^2-k_x^2)+C_1k_z^2+C_0-\tilde{\mu}_1-E\right]^2
-\left[M_2(p^2-k_x^2) +\gamma (k_z^2-m-m_1)\right]^2+v^2 (p^2-k_x^2) &=& 0 \quad \mbox{(up)},
\label{char-eq-up}\\
\left[ -C_2(p^2-k_x^2)+C_1k_z^2+C_0+\tilde{\mu}_1-E\right]^2
-\left[M_2(p^2-k_x^2) +\gamma (k_z^2-m+m_1)\right]^2+v^2 (p^2-k_x^2)&=& 0 \quad \mbox{(down)},
\label{char-eq-down}
\end{eqnarray}
cf. Eq.~(\ref{char-eq-no-alpha}). Then, the analysis of the surface Fermi arcs follows very closely
the analysis in Sec.~\ref{sec:2x2ModelC2M2}.

\begin{figure*}[ht!]
\begin{minipage}[ht]{0.245\linewidth}
\center{\includegraphics[width=1.0\linewidth]{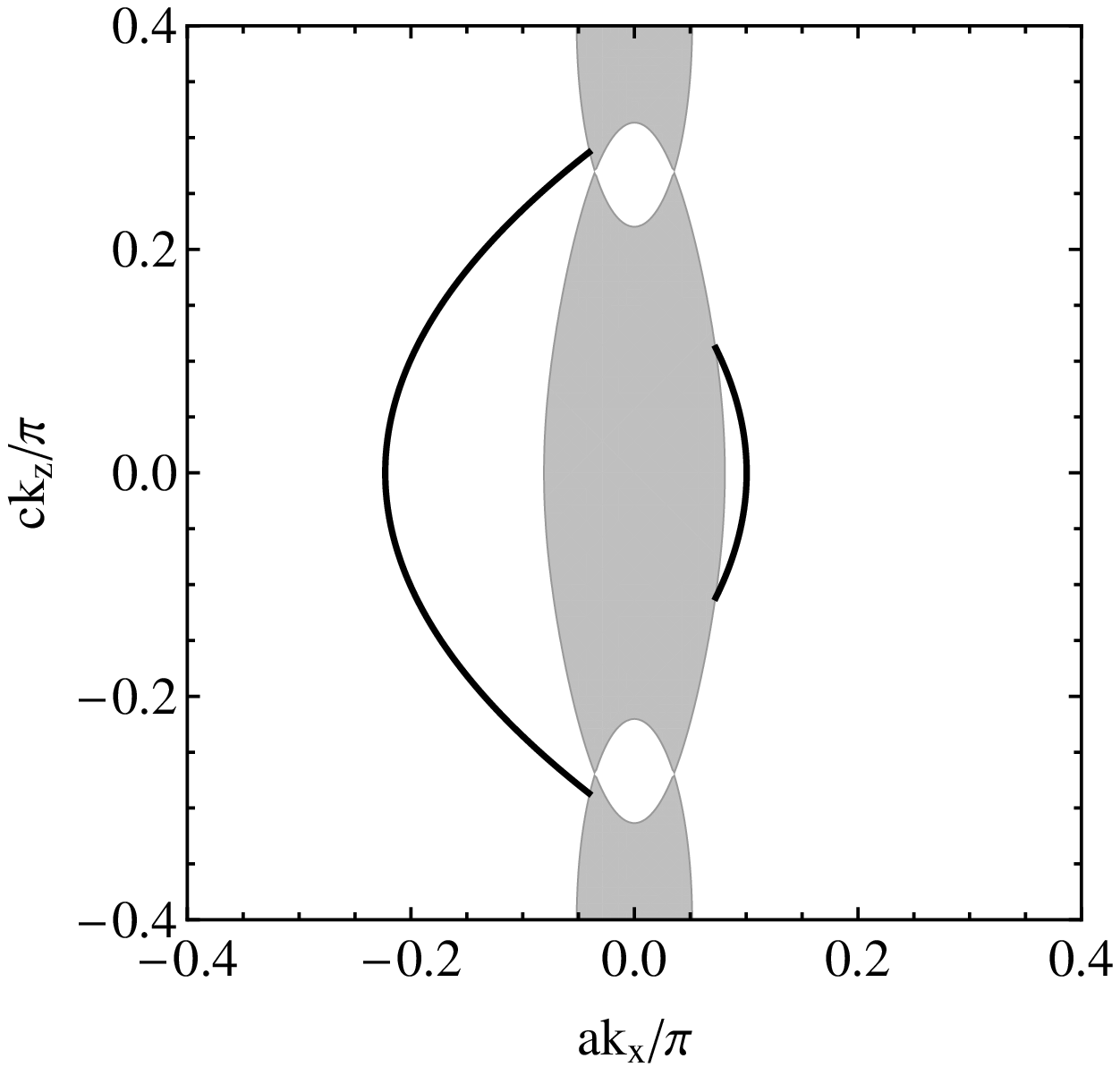} \\
{\small (a) $m_1=10^{-4}$, $\tilde{\mu}_1=0.05$}}
\end{minipage}
\begin{minipage}[ht]{0.245\linewidth}
\center{\includegraphics[width=1.0\linewidth]{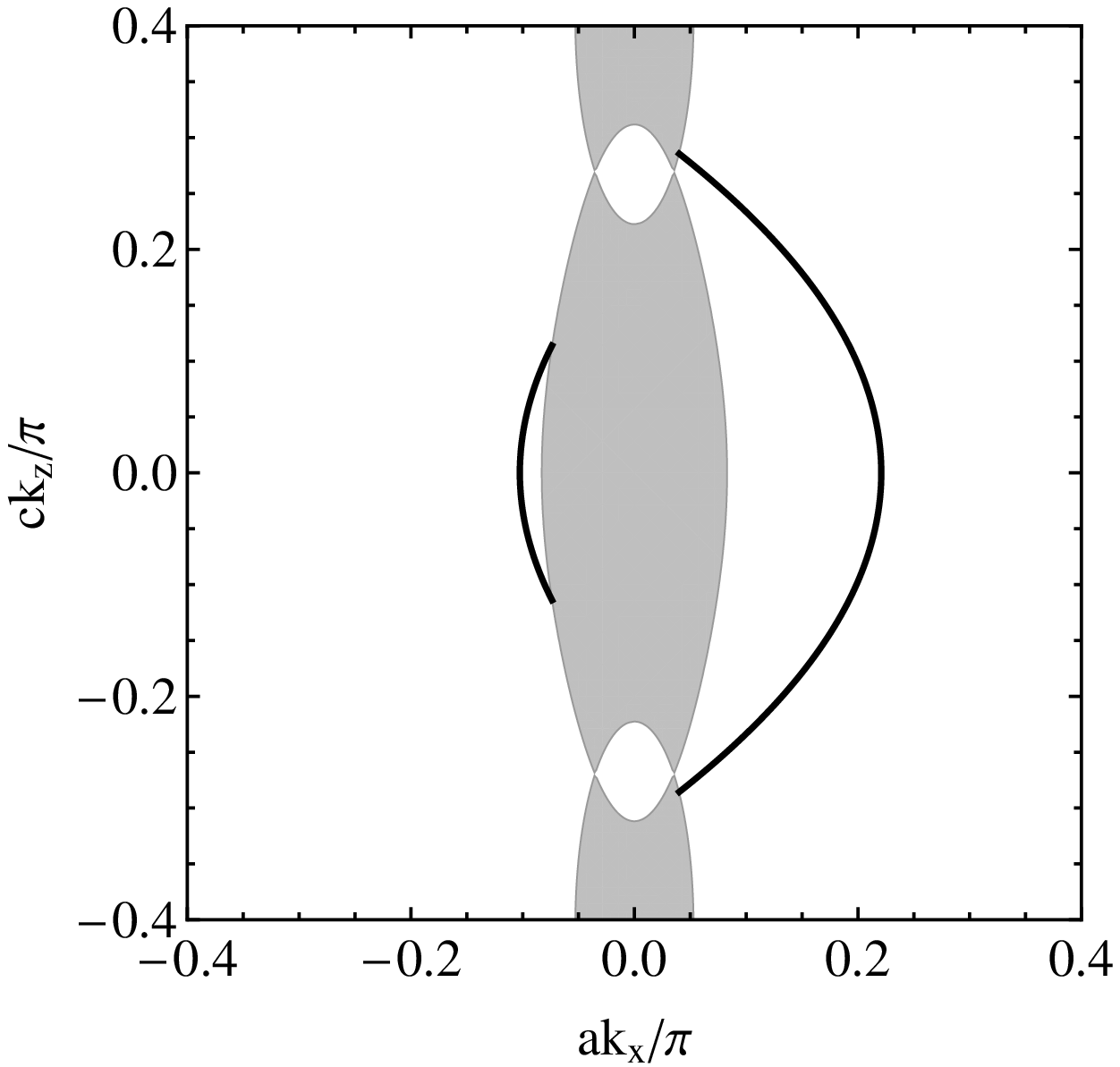} \\
{\small (b) $m_1=10^{-4}$, $\tilde{\mu}_1=-0.05$}}
\end{minipage}
\begin{minipage}[ht]{0.245\linewidth}
\center{\includegraphics[width=1.\linewidth]{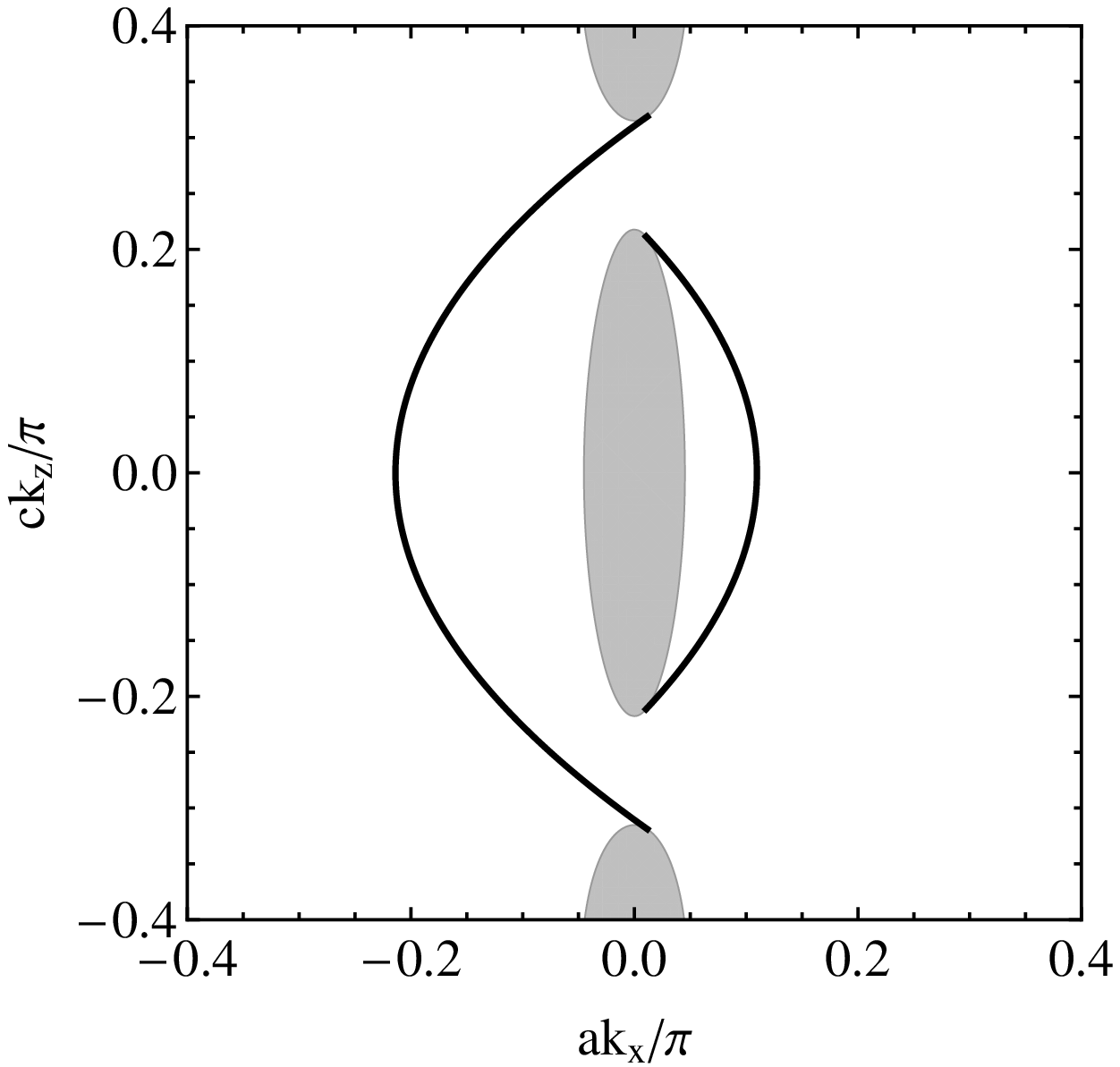} \\
{\small (c) $m_1=0.005$, $\tilde{\mu}_1=10^{-4}$}}
\end{minipage}
\begin{minipage}[ht]{0.245\linewidth}
\center{\includegraphics[width=1.0\linewidth]{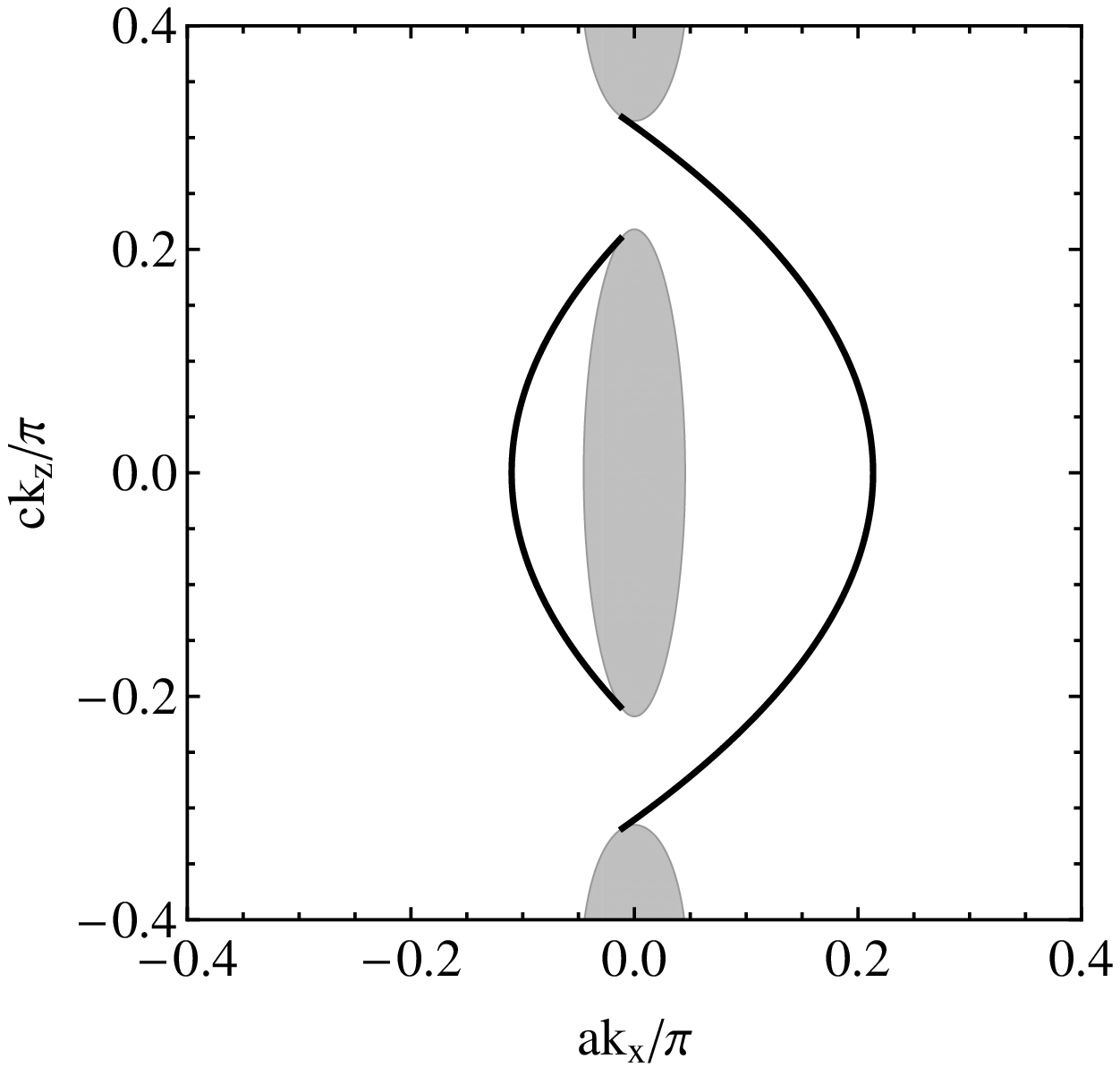} \\
{\small (d) $m_1=-0.005$, $\tilde{\mu}_1=10^{-4}$}}
\end{minipage}\\[5mm]
\begin{minipage}[ht]{0.245\linewidth}
\center{\includegraphics[width=1.0\linewidth]{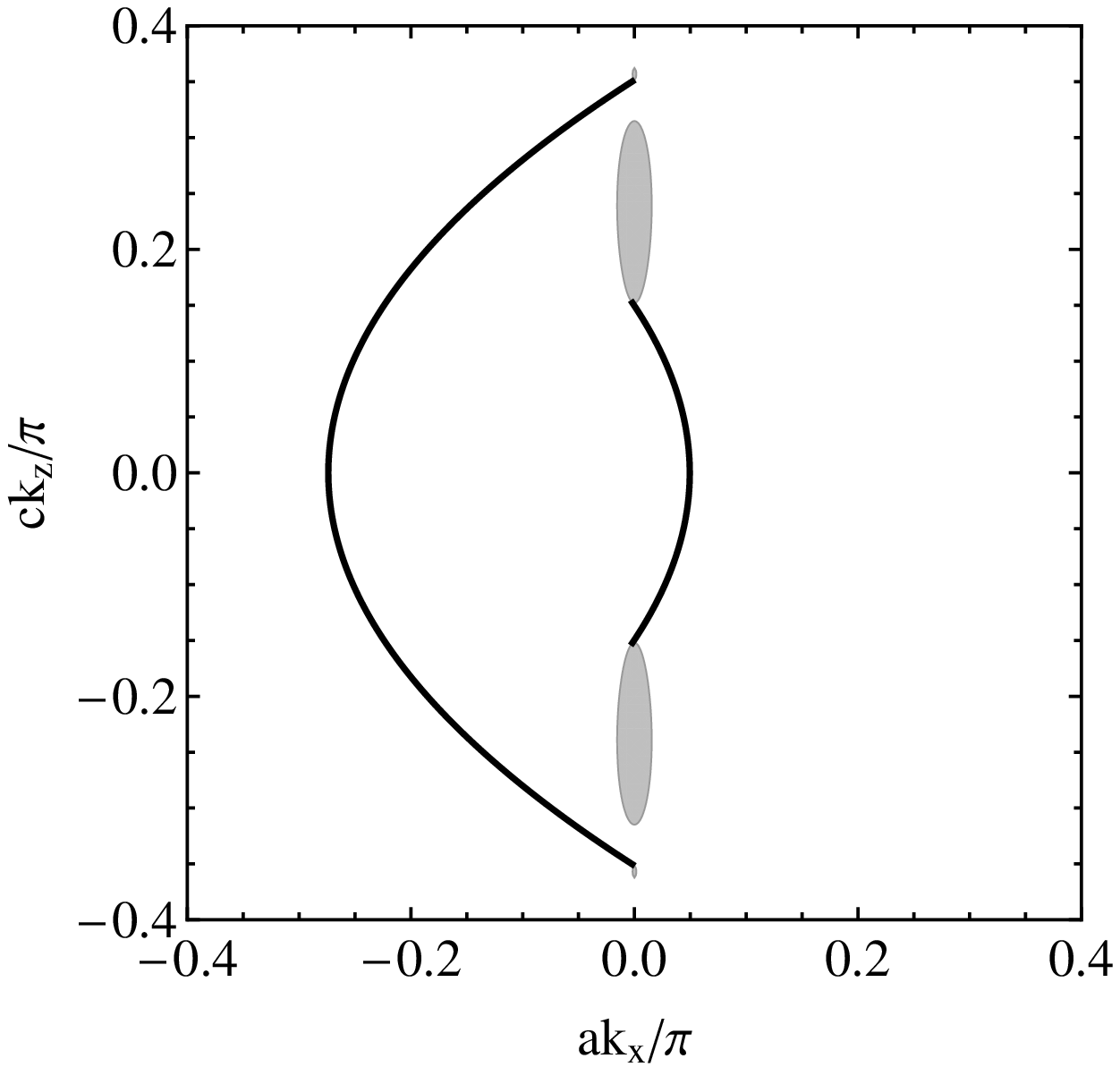} \\
{\small (e) $m_1=0.005$, $\tilde{\mu}_1=0.05$}}
\end{minipage}
\begin{minipage}[ht]{0.245\linewidth}
\center{\includegraphics[width=1.0\linewidth]{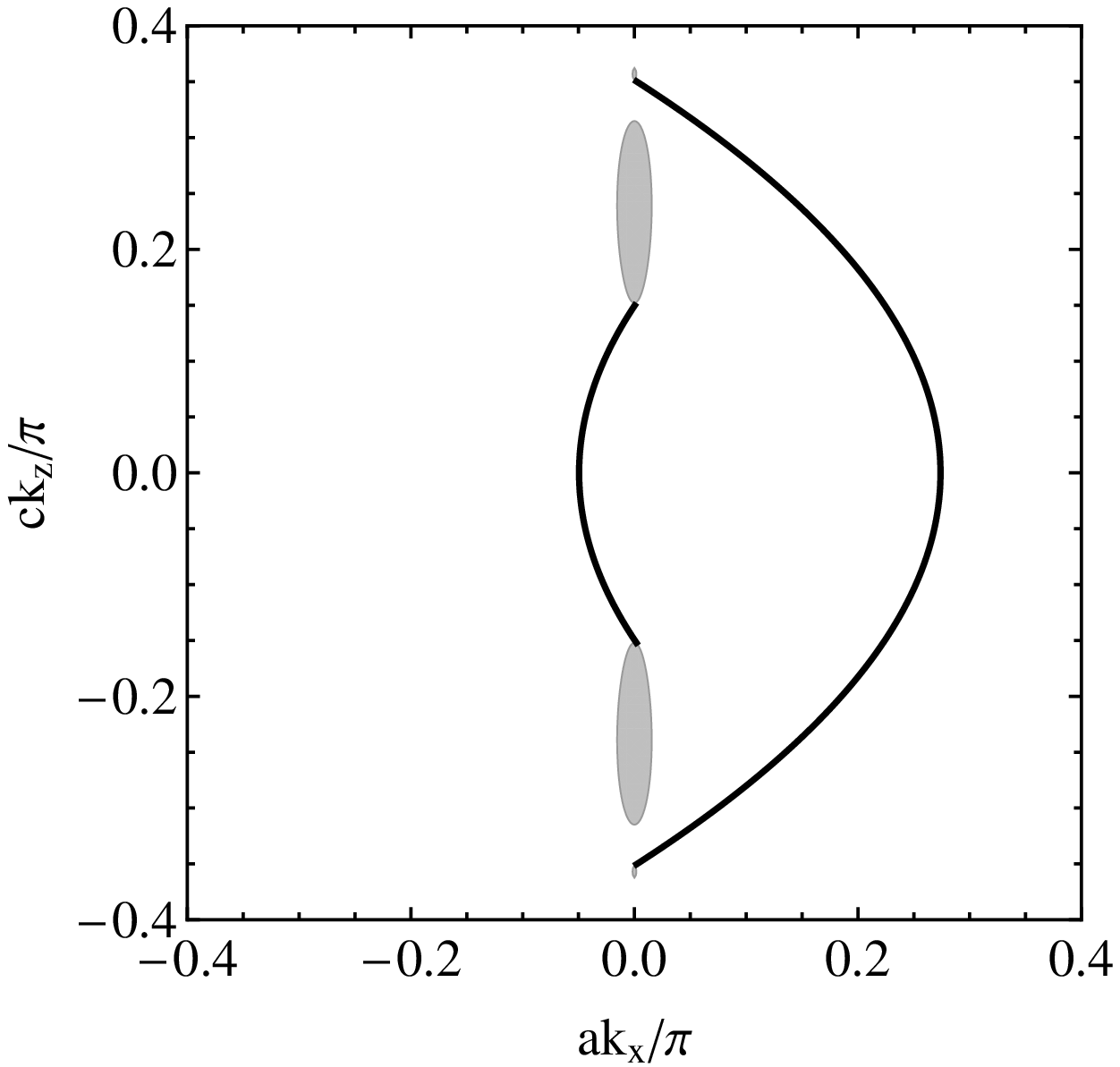} \\
{\small (f) $m_1=-0.005$, $\tilde{\mu}_1=-0.05$}}
\end{minipage}
\begin{minipage}[ht]{0.245\linewidth}
\center{\includegraphics[width=1.0\linewidth]{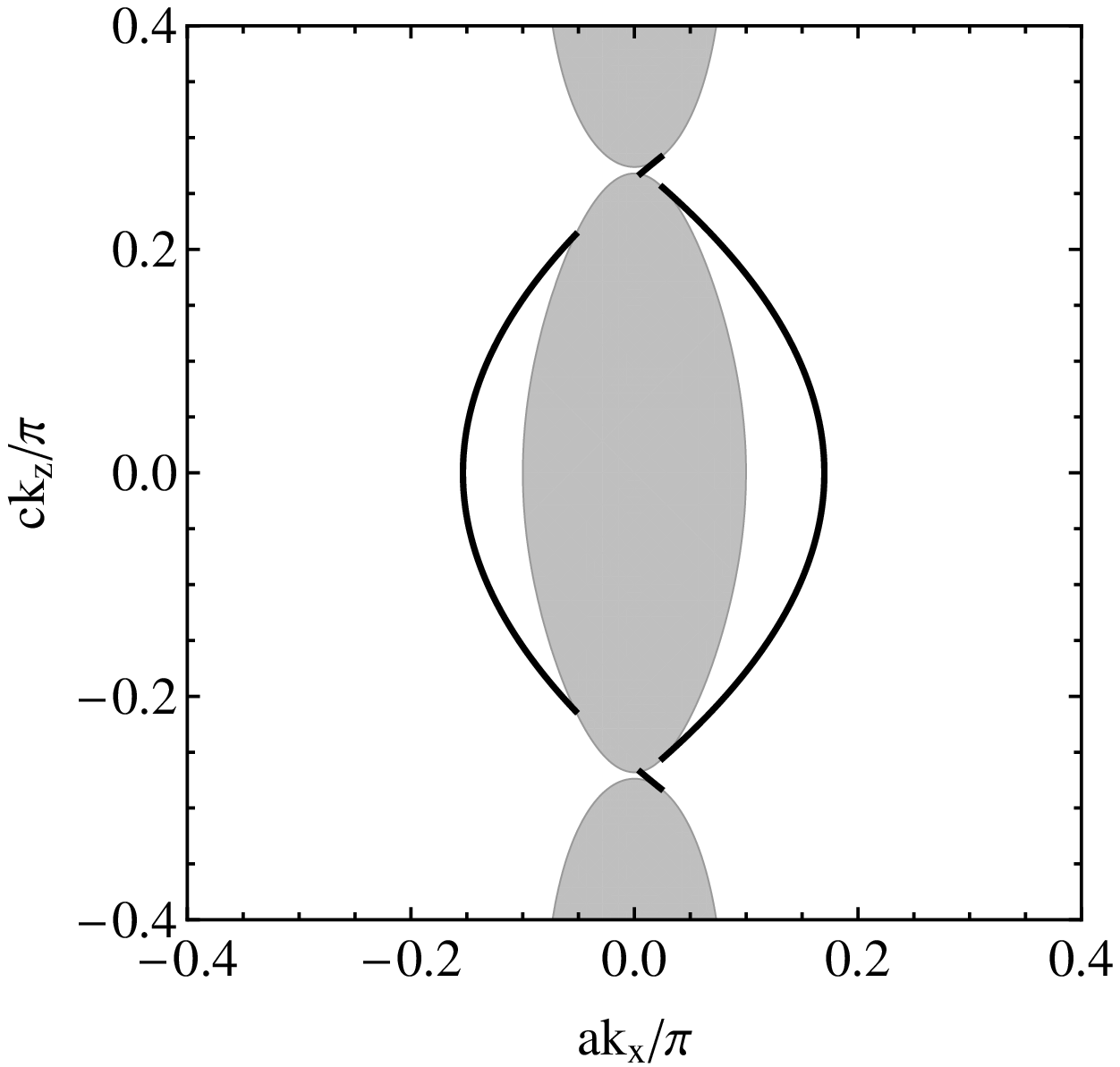} \\
{\small (g) $m_1=0.005$, $\tilde{\mu}_1=-0.05$}}
\end{minipage}
\begin{minipage}[ht]{0.245\linewidth}
\center{\includegraphics[width=1.0\linewidth]{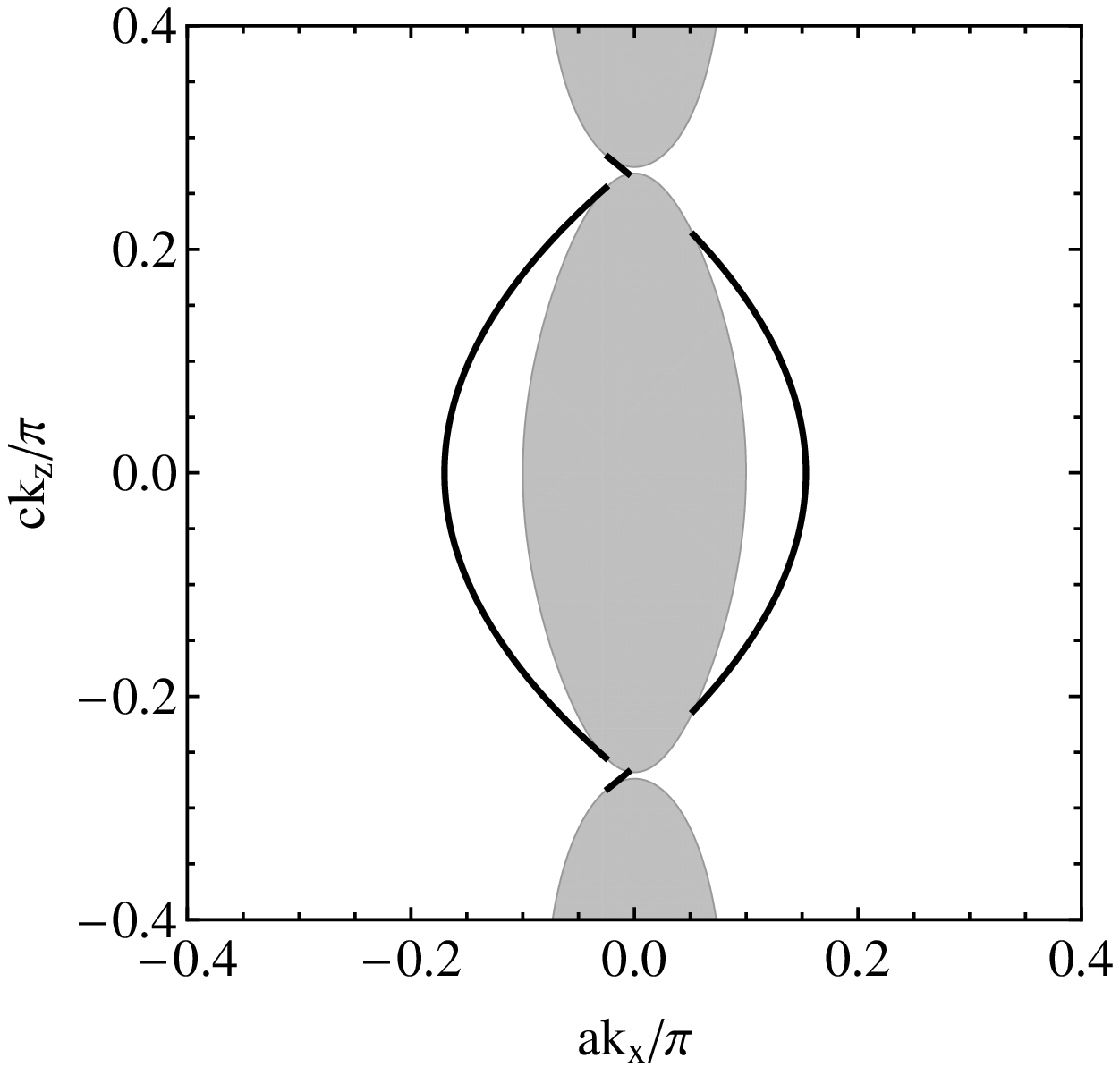} \\
{\small (h) $m_1=-0.005$, $\tilde{\mu}_1=0.05$}}
\end{minipage}
\caption{The Fermi arcs solutions (thick black lines) in the model with the symmetry breaking
parameters $m_1$ and $\tilde{\mu}_1$ at $E=0$. The shaded regions represent the projections
of the bulk Fermi surfaces onto the $k_xk_z$ plane. The values of $m_1$ and $\tilde{\mu}_1$
are given in units of $\mbox{\AA}^{-2}$ and $\mbox{eV}$, respectively.}
\label{fig:arcs_dynamical}
\end{figure*}

A number of representative numerical solutions for the Fermi surface arcs in the model with the
symmetry breaking parameters $m_1$ and $\tilde{\mu}_1$ are shown in Fig.~\ref{fig:arcs_dynamical}.
The results are obtained for the Fermi energy $E=0$. In order to shed light on the
origin of the individual arcs, in the same figure we also show the projections (shaded regions)
of the bulk Fermi surfaces onto the $k_xk_z$ plane. Such a representation reveals that
some of the Fermi arcs link {\it disconnected} sheets of the bulk Fermi surface \cite{Haldane},
while others link different points of the {\it same} bulk Fermi surface sheet.

As suggested by the physical meaning of the symmetry breaking parameters, $m_1$ and
$\tilde{\mu}_1$, the Fermi surface arcs for the up and down Weyl sectors of the theory are
not transformed into each other by a mirror symmetry. In addition to the expected effects of
(i) changing the length of the arcs (primarily due to nonzero $m_1$) and (ii) shifting the arcs'
position in the $k_x$ direction (primarily due to nonzero $\tilde{\mu}_1$), we also see some
qualitative changes in the shape and branching of the arcs. By comparing
Eqs.~(\ref{char-eq-up}) and (\ref{char-eq-down}) for the two sectors of the theory, we find
that the whole asymmetric sets of the Fermi arcs turn into their mirror reflections when
both parameters $m_1$ and $\tilde{\mu}_1$ change their signs. Examples of two pairs of such
mirror configurations are shown in panels (e)--(f) and (g)--(h) in Fig.~\ref{fig:arcs_dynamical}.
[Strictly speaking, the other two pairs of configurations, see (a)--(b) and (c)--(d), are not exact
mirror reflections of each other because one of the symmetry breaking parameters does not
change the sign. Because of a smallness of the parameter, there is an appearance
of approximate mirror configurations.]

It is interesting to point out that different topologies of the global (bulk-plus-arcs) Fermi
hypersurfaces, including the bulk sheets and the surface Fermi arcs, are possible. For example,
for a range of symmetry breaking parameters, represented by panels (c), (d), (e) and (f) in
Fig.~\ref{fig:arcs_dynamical}, we find that the global Fermi hypersurfaces consist of pairs of
clearly disconnected parts. This is in contrast to the configurations in panels (a) and (b), where
different parts touch at four points, and in contrast to the configurations in panels (g) and (h),
where all parts of the global Fermi hypersurfaces are linked by the Fermi arcs. If samples with
completely disconnected parts of the global Fermi hypersurfaces are indeed possible, they will
be very interesting to study in experiments.

As we see from panels (g) and (h) in Fig.~\ref{fig:arcs_dynamical}, there are also
qualitatively new types of the Fermi arcs possible for a range of symmetry breaking parameters.
In particular, we find a pair of ``short" branches of the Fermi arcs that split off from the usual
``long" arcs. To the best of our knowledge, the corresponding short arcs have not been predicted
before. So far, we could not establish a general criterion for the existence of the short arcs.
In the configurations in panels (g) and (h), they play a profound role by linking two disconnected
sheets of the bulk Fermi surface.

\section{Conclusion}
\label{Conclusion}

In this paper, we studied the surface Fermi arc states by employing a continuum low-energy effective
model. The use of analytical methods and a realistic low-energy model provide a deeper insight into
the physical properties and characteristics of the surface Fermi arcs. In particular, we were able to
classify the Fermi arcs with respect to the ud-parity and reconfirm the $\mathbb{Z}_2$ Weyl structure
of $\mathrm{A_3Bi}$ semimetals \cite{Gorbar:2014sja}. In this context, it should be noted
that the experimental observation of the corresponding Fermi arc states have been recently reported
for $\mathrm{Na_3Bi}$ \cite{science.1256742}. While in agreement with the claimed topological
semimetal structure, such an observation does not confirm it unambiguously. That is because the Fermi
arc states are also possible in Dirac materials where the $\mathbb{Z}_2$ Weyl structure is absent
\cite{WangWeng,Vishwanath}. The unambiguous confirmation of the $\mathbb{Z}_2$ Weyl structure could,
however, be established via the quantum oscillations, whose period should dependent on the thickness of
the semimetal in the same way as in true Weyl semimetals \cite{Vishwanath,Gorbar:2014qta}.

By introducing the effects of several possible symmetry breaking terms, we show that the $\mathbb{Z}_2$
Weyl structure of $\mathrm{A_3Bi}$ is destroyed in a very special way: the compounds become true
Weyl semimetals. We suggest that this finding can be tested in experiment. For example, by taking
into account that the mirror-symmetric pairs of surface Fermi arcs in clean $\mathrm{A_3Bi}$ get
distorted upon the introduction of explicit symmetry breaking (e.g., by magnetic doping), a number
of specific features (size, shape and number of branches) should be seen in the surface Fermi arcs.
The corresponding properties could be studied, for example, by analyzing the quantum oscillations
sensitive to the surface states of this type \cite{Vishwanath}. In the absence of symmetry
breaking, there will be a unique period of oscillations dependent in a specific way on the thickness of
the semimetal slab \cite{Gorbar:2014qta}. On the other hand, the breaking of symmetry will produce pairs
of inequivalent arcs of different lengths and the observation of two incommensurate periods of oscillations
will be expected. In principle, by making use of the analytical results in this study, the details of the
oscillations could be used to estimate the magnitude of the symmetry breaking terms.

\acknowledgments
The work of E.V.G. was supported partially by the Ukrainian State Foundation for Fundamental Research.
The work of V.A.M. was supported by the Natural Sciences and Engineering Research Council of Canada.
The work of I.A.S. was supported by the U.S. National Science Foundation under Grant No.~PHY-1404232.

\appendix

\section{Derivations of surface Fermi arcs solutions}
\label{AppExtra}

In this Appendix we present the key technical details of deriving the
surface Fermi arcs solutions in the $2\times 2$ model, introduced in
Sec.~\ref{sec:2x2ModelC2M2}, and in the $4\times 4$ model,
introduced in Sec.~\ref{sec:Realistic4x4Model}.

\subsection{Surface Fermi arcs in $2\times 2$ model}
\label{AppExtra1}

Let us start with the analysis of the surface Fermi arc states in the $2\times 2$ model, introduced
in Sec.~\ref{sec:2x2ModelC2M2}. The problem reduces to solving the eigenvalues problem given
by Eqs.~(\ref{psi-equation-1-01}) and (\ref{psi-equation-2-01}) at $y>0$ (semimetal), as well as
a similar set of equations at $y<0$ (vacuum), but $m$ replaced by $-\tilde{m}$. The corresponding
set of equations should be also supplemented by the boundary conditions at the vacuum-semimetal
interface, i.e.,
\begin{eqnarray}
\tilde{\psi}_1(-0) &=& \psi_1(+0), \label{cross-liking0}\\
\tilde{\psi}_2(-0) &=& \psi_2(+0), \label{cross-liking1}\\
-C_2\partial_y\tilde{\psi}_1(-0)+M_2\partial_y\tilde{\psi}_2(-0) &=& -C_2\partial_y\psi_1(+0)+M_2\partial_y\psi_2(+0),  \label{cross-liking2}\\
-M_2\partial_y\tilde{\psi}_1(-0)+C_2\partial_y\tilde{\psi}_2(-0) &=& -M_2\partial_y\psi_1(+0)+C_2\partial_y\psi_2(+0), \label{cross-liking3}
\end{eqnarray}
where $\tilde{\psi}_{1, 2}(y)$ correspond to the vacuum region at $y<0$.

Inside the semimetal ($y>0$), the surface state solutions should have the following form:
\begin{equation}
\Psi_{y>0} (y) =\left(\begin{array}{r}
a \\
b
\end{array}\right)e^{- p y}.
\end{equation}
By substituting this ansatz in Eqs.~(\ref{psi-equation-1-01}) and (\ref{psi-equation-2-01}), we arrive
at the following set of linear equations for the spinor components $a$ and $b$:
\begin{eqnarray}
\label{psi-equation-1-01a}
\left[ C_2(k_x^2 - p^2)-v k_x+C_1k_z^2+C_0-E\right]a
+ \left[-M_2(k_x^2 - p^2)-v p +\gamma k_z^2-\gamma m\right]b &=&  0, \\
\label{psi-equation-2-01b}
 \left[-M_2(k_x^2 - p^2)+v p +\gamma k_z^2-\gamma m\right]a
+\left[ C_2(k_x^2 - p^2)+v k_x+C_1k_z^2+C_0-E\right]b &=&  0.
\end{eqnarray}
A nontrivial solution exists when the following characteristic equation is satisfied:
\begin{equation}
\left[ -C_2(p^2-k_x^2)+C_1k_z^2+C_0-E\right]^2
-\left[M_2(p^2-k_x^2) +\gamma k_z^2-\gamma m\right]^2
+v^2 (p^2-k_x^2)=0 .
\label{char-eq-no-alpha}
\end{equation}
The solutions to this equation are $p=\pm p_1$ and $p=\pm p_2$, where
\begin{equation}
p_1 = \sqrt{k_x^2-\frac{X+ \sqrt{X^2+Y}}{2(M_2^2-C_2^2)}},
\qquad \qquad
p_2 = \sqrt{k_x^2-\frac{X- \sqrt{X^2+Y}}{2(M_2^2-C_2^2)}}.
\label{p1p2-two-component}
\end{equation}
Here we introduced the following shorthand notations:
\begin{eqnarray}
X &\equiv & 2C_2(C_1k_z^2+C_0-E)+2\gamma M_2(k_z^2- m)-v^2,
\label{X-two-component} \\
Y &\equiv & 4(M_2^2-C_2^2)\left[(C_1k_z^2+C_0-E)^2-\gamma^2 (k_z^2-m)^2\right].
\label{Y-two-component}
\end{eqnarray}
The spinor components $a$ and $b$ of the corresponding nontrivial solution satisfy the
constraint
\begin{equation}
\frac{b}{a} = Q(p,k_x) \equiv \frac{-C_2(p^2-k_x^2)+C_1k_z^2+C_0-E-v k_x}{-M_2(p^2-k_x^2)-\gamma (k_z^2-m)+vp}.
\label{def-Qpkx}
\end{equation}
Inside the semimetal ($y>0$), the wave function should fall off with increasing $y$. Thus, we use
only the negative exponents in the general solution, i.e.,
\begin{equation}
\Psi_{y>0}(y) \simeq  a_1 \left(\begin{array}{r}
1 \\
Q_1
\end{array}\right)e^{-p_1 y}
+a_2 \left(\begin{array}{l}
1 \\
Q_2
\end{array}\right)e^{-p_2 y},
\end{equation}
where $Q_i \equiv Q(p_i,k_x)$ with $i=1,2$.

In order to find the vacuum solution ($y<0$), we replace $m\rightarrow-\tilde{m}$ and take the
limit $\tilde{m}\to \infty$. This leads to the following general solution on the vacuum side:
\begin{equation}
\Psi_{y<0}(y) \simeq  \frac{\tilde{a}_1}{\sqrt{\gamma \tilde{m}}} \left(\begin{array}{r}
1 \\
-1
\end{array}\right)e^{\tilde{p}_1 y}
+\frac{\tilde{a}_2}{\sqrt{\gamma \tilde{m}}} \left(\begin{array}{l}
1 \\
1
\end{array}\right)e^{\tilde{p}_2 y} ,
\label{vac_solution1}
\end{equation}
where, for convenience, we took the overall constants to be inversely proportional to
$\sqrt{\gamma \tilde{m}}$. The exponents in the vacuum solution are determined by
\begin{equation}
\tilde{p}_1\simeq \sqrt{\frac{\gamma \tilde{m} }{-M_2- C_2}},
\qquad
\tilde{p}_2\simeq \sqrt{\frac{\gamma \tilde{m} }{-M_2+ C_2}} .
\end{equation}
It is interesting to note that the conditions of the wave function continuity, see Eqs.~(\ref{cross-liking0})
and (\ref{cross-liking1}), are the only important conditions to be satisfied. Indeed, the nontrivial solution
of Eq.~(\ref{cross-liking0}) in the limit $\tilde{m}\to \infty$ implies that $a_1=-a_2\neq 0$. This is consistent
with Eq.~(\ref{cross-liking1}) only when $Q_1= Q_2$.
Concerning the remaining boundary conditions in Eqs.~(\ref{cross-liking2})
and (\ref{cross-liking3}), enforcing the continuity of the wave function derivative, they do
not add any additional constraints. In fact, they are needed only for determining the vacuum
spinor components $\tilde{a}_1$ and $\tilde{a}_2$ in terms of the nontrivial components
$a_1$ and $a_2$ in the semimetal. Such (finite) solutions always exist. However,
as is clear from Eq.~(\ref{vac_solution1}), the vacuum solution have no much physical content
because it vanishes in the limit $\tilde{m}\to \infty$.

In conclusion, the boundary conditions at $y=0$ are satisfied and, therefore, a nontrivial solution
exists when $Q_1= Q_2$. The explicit form of the corresponding condition is given in
Eq.~(\ref{Q1=Q2_alpha=0}) in the main text.

\subsection{Surface Fermi arcs in $4\times 4$ model}
\label{AppExtra2}

The analysis of the realistic $4\times 4$ model introduced in Sec.~\ref{sec:Realistic4x4Model} is
slightly more involved, but qualitatively similar. The eigenvalues problem in this case is given by
Eqs.~(\ref{psi-equation-1-four}) through (\ref{psi-equation-4-four}) at $y>0$ (semimetal), as well as
a similar set of equations at $y<0$ (vacuum), but with $m$ replaced by $-\tilde{m}$. The conditions of
continuity of the wave functions and their derivatives across the vacuum-semimetal
surface at $y=0$ are given by
\begin{eqnarray}
\tilde{\psi}_i(-0) &=& \psi_i(+0), \quad \mbox{for}\quad i=1,2,3,4,
\label{cross-liking-four1} \\
-C_2\partial_y\tilde{\psi}_1(-0)+M_2\partial_y\tilde{\psi}_2(-0)-\alpha k_z \partial_y \tilde{\psi}_3(-0)
&=& -C_2\partial_y\psi_1(+0)+M_2\partial_y\psi_2(+0)-\alpha k_z \partial_y \psi_3(+0),  \label{cross-liking-four2} \\
M_2\partial_y\tilde{\psi}_1(-0)-C_2\partial_y\tilde{\psi}_2(-0)+ \alpha k_z \partial_y \tilde{\psi}_4(-0)
&=& M_2\partial_y\psi_1(+0) -C_2\partial_y\psi_2(+0)+\alpha k_z \partial_y \psi_4(+0),
\label{cross-liking-four3} \\
-\alpha k_z \partial_y \tilde{\psi}_1(-0)-C_2\partial_y\tilde{\psi}_3(-0)+M_2\partial_y\tilde{\psi}_4(-0)
&=& -\alpha k_z \partial_y \psi_1(+0) -C_2\partial_y\psi_3(+0)+M_2\partial_y\psi_4(+0),
\label{cross-liking-four4} \\
\alpha k_z \partial_y \tilde{\psi}_2(-0) +M_2\partial_y\tilde{\psi}_3(-0)-C_2\partial_y\tilde{\psi}_4(-0)
&=& \alpha k_z \partial_y \psi_2(+0) +M_2\partial_y\psi_3(+0)-C_2\partial_y\psi_4(+0).
\label{cross-liking-four5}
\end{eqnarray}
In the semimetal ($y>0$), we look for a general surface state solution in the form
\begin{equation}
\Psi_{y>0} (y) =\left(\begin{array}{r}
a \\
b\\
c\\
d
\end{array}\right)e^{- p y}.
\end{equation}
A nontrivial solution of this type exists when $p$ is a solution to the
following characteristic equation:
\begin{equation}
\left[ -C_2(p^2-k_x^2)+C_1k_z^2+C_0-E\right]^2
-\left[M_2(p^2-k_x^2) +\gamma k_z^2-\gamma m\right]^2
+v^2 (p^2-k_x^2)-\alpha^2 k_z^2(p^2-k_x^2)^2=0 .
\label{char-eq-1}
\end{equation}
(Strictly speaking, the characteristic equation has the square on the left hand side, implying that the
degeneracy of its solutions should be doubled.) This equation has {\em two} pairs of distinct solutions:
$p=\pm p_1$ and $p=\pm p_2$, where
\begin{equation}
p_{1,2}= \sqrt{k_x^2-\frac{X
\pm \sqrt{X^2+Y}}{2(M_2^2-C_2^2+\alpha^2 k_z^2)}} ,
\label{p12-four-comp}
\end{equation}
cf. Eq.~(\ref{p1p2-two-component}). Here, the expression for $X$ is the same as in the
$2\times 2$ model ($\alpha=0$) in Eq.~(\ref{X-two-component}), but the expression for
$Y$ is slightly different, i.e.,
\begin{equation}
Y \equiv 4(M_2^2-C_2^2+\alpha^2 k_z^2)\left[(C_1k_z^2+C_0-E)^2-\gamma^2 (k_z^2-m)^2\right].
\label{Y-four-component}
\end{equation}
Therefore, in the half-space occupied by the semimetal ($y>0$), the wave function should have the
following general form:
\begin{equation}
\Psi_{y>0}(y) \simeq  \left(\begin{array}{c}
a_1  \\
Q_1^{+} a_1 - T_1^{+} c_1  \\
c_1  \\
-T_1^{-} a_1 + Q_1^{-} c_1
\end{array}\right)e^{-p_1 y}
+ \left(\begin{array}{c}
a_2  \\
Q_2^{+} a_2 - T_2^{+} c_2  \\
c_2  \\
-T_2^{-} a_2 + Q_2^{-} c_2
\end{array}\right)e^{-p_2 y} ,
\label{sol-bd-ac}
\end{equation}
where we  also introduced the shorthand notation: $Q_{i}^{\pm} \equiv Q(p_i,\pm k_x)$
and $T_{i}^{\pm} \equiv T(p_i,\pm k_x)$. Here the function $Q(p,k_x)$ is the same as in
Eq.~(\ref{def-Qpkx}) and
\begin{equation}
T(p,k_x)= \frac{\alpha k_z (p+k_x)^2} {-M_2(p^2-k_x^2)-\gamma (k_z^2-m)+vp}.
\label{def-Tpkx}
\end{equation}

In the vacuum solution ($y<0$), we replace $m\rightarrow-\tilde{m}$ and take the limit $\tilde{m}\to \infty$.
In this case, a simple analysis leads to the following solution:
\begin{equation}
\Psi_{y<0}(y) \simeq  \frac{1}{\sqrt{\gamma \tilde{m}}} \left(\begin{array}{c}
\tilde{a}_1 \\
\frac{\tilde{a}_1 C_2+\alpha k_z \tilde{c}_1}{\sqrt{C_2^2-\alpha^2 k_z^2}}\\
\tilde{c}_1 \\
\frac{\alpha k_z \tilde{a}_1 + C_2\tilde{c}_1}{\sqrt{C_2^2-\alpha^2 k_z^2}}
\end{array}\right)e^{\tilde{p}_1 y}
+\frac{1}{\sqrt{\gamma \tilde{m}}} \left(\begin{array}{c}
\tilde{a}_2 \\
-\frac{\tilde{a}_2 C_2+\alpha k_z \tilde{c}_2}{\sqrt{C_2^2-\alpha^2 k_z^2}}\\
\tilde{c}_2 \\
-\frac{\alpha k_z \tilde{a}_2 + C_2\tilde{c}_2}{\sqrt{C_2^2-\alpha^2 k_z^2}}
\end{array}\right)e^{\tilde{p}_2 y},
\label{vac-solution-four-comp}
\end{equation}
where, for convenience, we introduced an overall constant inversely proportional
to $\sqrt{\gamma \tilde{m}}$. The explicit form of the exponents in this solution is
determined by
\begin{equation}
\tilde{p}_1\simeq \sqrt{\frac{\gamma \tilde{m} }{-M_2+ \sqrt{C_2^2-\alpha^2 k_z^2}}},
\qquad
\tilde{p}_2\simeq \sqrt{\frac{\gamma \tilde{m} }{-M_2- \sqrt{C_2^2-\alpha^2 k_z^2}}} .
\end{equation}
Note that the signs in the exponents of the vacuum solution (\ref{vac-solution-four-comp})
are chosen so that the wave function vanishes at $y\to -\infty$.

The conditions of the continuity of the wave function in Eq.~(\ref{cross-liking-four1}) lead
to the following constraints:
\begin{eqnarray}
0 &=& Q_1^{+}  a_1 - T_1^{+} c_1+Q_2^{+} a_2 - T_2^{+}c_2,
\label{first-four-conditions-bd1}
\\
0 &=& T_1^{-} a_1 - Q_1^{-} c_1+T_2^{-} a_2 - Q_2^{-} c_2,
\label{first-four-conditions-bd2}
\end{eqnarray}
together with $a_2=-a_1$ and $c_2=-c_1$. Here, we took into account that the left hand
side of Eq.~(\ref{cross-liking-four1}) vanishes in the limit $\tilde{m} \to \infty$.

As in the case of a two-component model, discussed in Appendix~\ref{AppExtra1}, there is
no need to satisfy the continuity conditions for the wave function derivatives, given by
Eqs.~(\ref{cross-liking-four2}) through (\ref{cross-liking-four5}). The reason is that these
conditions add no additional constraints on the spinor solutions in the semimetal.
They are needed only for determining the components of the vacuum solution at $y<0$.
Since the latter has no physical content in the limit $\tilde{m} \to \infty$, we can safely ignore
the conditions in Eqs.~(\ref{cross-liking-four2}) through (\ref{cross-liking-four5}).

In order to have a nontrivial solution to Eqs.~(\ref{first-four-conditions-bd1}) and (\ref{first-four-conditions-bd2}),
the condition in Eq.~(\ref{key-equation}) in the main text should be satisfied.

\section{Symmetries and surface Fermi arcs bispinors}
\label{AppA}

In this Appendix, we discuss the properties of the Fermi arc surface states with respect
to the discrete symmetries $U_{\chi}\equiv U\Pi_{k_z}$ and $\tilde{U}\Pi_{k_x}$, introduced in
Sec.~\ref{sec:symmetries}. To start with, let us note that the general Fermi arc spinor in
Eq.~(\ref{sol-bd-ac}) contains all possible solutions. It is possible to classify these solutions
with respect to the discrete symmetry $U_{\chi}$ by choosing them as eigenstates of the
operator $U_{\chi}$.

In order to construct the first group of solutions, we use the relations in Eqs.~(\ref{first-four-conditions-bd1})
and (\ref{first-four-conditions-bd2}) and rewrite the Fermi arc spinor in Eq.~(\ref{sol-bd-ac}) in the
following form:
\begin{equation}
\Psi_{+}=
a_1\left(\begin{array}{c}
1 \\
Q_1^{+}-T_1^{+}\frac{T_1^{-}-T_2^{-}}{Q_1^{-}-Q_2^{-}} \\
\frac{T_1^{-}-T_2^{-}}{Q_1^{-}-Q_2^{-}} \\
-T_1^{-}+Q_1^{-}\frac{T_1^{-}-T_2^{-}}{Q_1^{-}-Q_2^{-}}
\end{array}\right)e^{- p_1 y}
- a_1\left(\begin{array}{c}
1 \\
Q_2^{+}-T_2^{+}\frac{T_1^{-}-T_2^{-}}{Q_1^{-}-Q_2^{-}} \\
\frac{T_1^{-}-T_2^{-}}{Q_1^{-}-Q_2^{-}} \\
-T_2^{-}+Q_2^{-}\frac{T_1^{-}-T_2^{-}}{Q_1^{-}-Q_2^{-}}
\end{array}\right)e^{- p_2 y} .
\label{psi-all}
\end{equation}
By noting that $p_i$'s (with $i=1,2$), defined in Eq.~(\ref{p12-four-comp}), contain only
quadratic terms in momenta $k_x$ and $k_z$, we conclude that both of them are invariant
under the $\Pi_{k_z}$ and $\Pi_{k_x}$ transformations. The other quantities, used in
Eq.~(\ref{psi-all}), transform as follows:
\begin{eqnarray}
&\Pi_{k_z}Q_{i}^{\pm}=Q_{i}^{\pm}, \qquad &\Pi_{k_z}T_{i}^{\pm}=-T_{i}^{\pm}, \\
&\Pi_{k_x}Q_{i}^{\pm}=Q_{i}^{\mp}, \qquad &\Pi_{k_x}T_{i}^{\pm}=T_{i}^{\mp}.
\label{Q-T-transform}
\end{eqnarray}
It is straightforward to check that the spinor in Eq.~(\ref{psi-all}) is an eigenstate of the
operator $U_{\chi}$ with the eigenvalue $\chi=+1$. Indeed, by making use of the
definition of the matrix $U$, we find that
\begin{equation}
U\Psi_{+}=a_1\left(\begin{array}{c}
1 \\
Q_1^{+}-T_1^{+}\frac{T_1^{-}-T_2^{-}}{Q_1^{-}-Q_2^{-}} \\
-\left[\frac{T_1^{-}-T_2^{-}}{Q_1^{-}-Q_2^{-}}\right] \\
-\left[-T_1^{-}+Q_1^{-}\frac{T_1^{-}-T_2^{-}}{Q_1^{-}-Q_2^{-}}\right]
\end{array}\right)e^{- p_1 y} - a_1\left(\begin{array}{c}
1 \\
Q_2^{+}-T_2^{+}\frac{T_1^{-}-T_2^{-}}{Q_1^{-}-Q_2^{-}} \\
-\left[\frac{T_1^{-}-T_2^{-}}{Q_1^{-}-Q_2^{-}}\right] \\
-\left[-T_2^{-}+Q_2^{-}\frac{T_1^{-}-T_2^{-}}{Q_1^{-}-Q_2^{-}}\right]
\end{array}\right)e^{- p_2 y} .
\end{equation}
Then, by taking into account that $U_{\chi}\equiv U\Pi_{k_z}$, we see that
$U_{\chi} \Psi_{+}=\Psi_{+}$, as claimed.

By using the relations in Eqs.~(\ref{first-four-conditions-bd1}) and (\ref{first-four-conditions-bd2}),
the Fermi arc spinor in Eq.~(\ref{sol-bd-ac}) can be also rewritten in the following alternative form:
\begin{equation}
\Psi_{-}=c_1\left(\begin{array}{c}
\frac{T_1^{+}-T_2^{+}}{Q_1^{+}-Q_2^{+}} \\
Q_1^{+}\frac{T_1^{+}-T_2^{+}}{Q_1^{+}-Q_2^{+}}-T_1^{+} \\
1 \\
-T_1^{-}\frac{T_1^{+}-T_2^{+}}{Q_1^{+}-Q_2^{+}}+Q_1^{-}
\end{array}\right)e^{- p_1 y} - c_1\left(\begin{array}{c}
\frac{T_1^{+}-T_2^{+}}{Q_1^{+}-Q_2^{+}} \\
Q_2^{+}\frac{T_1^{+}-T_2^{+}}{Q_1^{+}-Q_2^{+}}-T_2^{+} \\
1 \\
-T_2^{-}\frac{T_1^{+}-T_2^{+}}{Q_1^{+}-Q_2^{+}}+Q_2^{-}
\end{array}\right)e^{- p_2 y} .
\label{psi-all-c1}
\end{equation}
In this case, as is easy to check, the spinor is an eigenstate of the operator $U_{\chi}$
with eigenvalue $\chi=-1$. Indeed, by making use of the definition of the matrix $U$,
we find that
\begin{equation}
U\Psi_{-}=c_1\left(\begin{array}{c}
\frac{T_1^{+}-T_2^{+}}{Q_1^{+}-Q_2^{+}} \\
Q_1^{+}\frac{T_1^{+}-T_2^{+}}{Q_1^{+}-Q_2^{+}}-T_1^{+} \\
-1 \\
-\left[-T_1^{-}\frac{T_1^{+}-T_2^{+}}{Q_1^{+}-Q_2^{+}}+Q_1^{-}\right]
\end{array}\right)e^{- p_1 y} - c_1\left(\begin{array}{c}
\frac{T_1^{+}-T_2^{+}}{Q_1^{+}-Q_2^{+}} \\
Q_2^{+}\frac{T_1^{+}-T_2^{+}}{Q_1^{+}-Q_2^{+}}-T_2^{+} \\
-1 \\
-\left[-T_2^{-}\frac{T_1^{+}-T_2^{+}}{Q_1^{+}-Q_2^{+}}+Q_2^{-}\right]
\end{array}\right)e^{- p_2 y},
\label{psi-Ukz-c1}
\end{equation}
which implies that $U_{\chi} \Psi_{-}=-\Psi_{-}$. In other words, the eigenstate $\Psi_{-}$ corresponds
to $\chi=-1$, as claimed.

Now, let us explore the implications of the $\tilde{U}\Pi_{k_x}$ symmetry in the model at hand.
By applying the corresponding operator to the eigenstates $\Psi_{\pm}$, we arrive at the
following results:
\begin{eqnarray}
\tilde{U}\Pi_{k_x} \Psi_{+} &=& a_1\left(\begin{array}{c}
\frac{Q_1^{-}-Q_2^{-}}{T_1^{-}-T_2^{-}} \\
-T_1^{+}+Q_1^{+}\frac{Q_1^{-}-Q_2^{-}}{T_1^{-}-T_2^{-}} \\
1 \\
Q_1^{-}-T_1^{-}\frac{Q_1^{-}-Q_2^{-}}{T_1^{-}-T_2^{-}}
\end{array}\right)e^{- p_1 y} - a_1\left(\begin{array}{c}
\frac{Q_1^{-}-Q_2^{-}}{T_1^{-}-T_2^{-}} \\
-T_2^{+}+Q_2^{+}\frac{Q_1^{-}-Q_2^{-}}{T_1^{-}-T_2^{-}} \\
1 \\
Q_2^{-}-T_2^{-}\frac{Q_1^{-}-Q_2^{-}}{T_1^{-}-T_2^{-}}
\end{array}\right)e^{- p_2 y} ,
\label{psi-Ukx} \\
\tilde{U}\Pi_{k_x} \Psi_{-} &=& c_1\left(\begin{array}{c}
1 \\
-T_1^{+}\frac{T_1^{-}-T_2^{-}}{Q_1^{-}-Q_2^{-}}+Q_1^{+} \\
\frac{T_1^{-}-T_2^{-}}{Q_1^{-}-Q_2^{-}} \\
Q_1^{-}\frac{T_1^{-}-T_2^{-}}{Q_1^{-}-Q_2^{-}}-T_1^{-}
\end{array}\right)e^{- p_1 y} - c_1\left(\begin{array}{c}
1 \\
-T_2^{+}\frac{T_1^{-}-T_2^{-}}{Q_1^{-}-Q_2^{-}}+Q_2^{+} \\
\frac{T_1^{-}-T_2^{-}}{Q_1^{-}-Q_2^{-}} \\
Q_2^{-}\frac{T_1^{-}-T_2^{-}}{Q_1^{-}-Q_2^{-}}-T_2^{-}
\end{array}\right)e^{- p_2 y}.
\label{psi-Ukx-c1}
\end{eqnarray}
These results show that $\Psi_{\pm}$ are not eigenstates of the operator $\tilde{U}\Pi_{k_x}$.
However, by taking into account the constraint in Eq.~(\ref{key-equation}), one can check
that the operator $\tilde{U}\Pi_{k_x}$ interchanges the two types of the states, i.e.,
$\Psi_{+}\leftrightarrow\Psi_{-}$.

In conclusion, the results of this Appendix confirm the claim in the main text of the paper
concerning the symmetry properties of the low-energy theory for $\mathrm{A_3Bi}$ (A=Na, K, Rb)
semimetals, as well as the classification of their Fermi arc states. These are in complete agreement
with the claim that the corresponding compounds are $\mathbb{Z}_2$ Weyl semimetals.


\begin{thebibliography}{99}

\bibitem{Geim}
K.~S.~Novoselov, A.~K.~Geim, S.~V.~Morozov, D.~Jiang, Y.~Zhang, S.~V.~Dubonos, I.~V.~Grigorieva, and A.~A.~Firsov,
Science {\bf 306}, 666 (2004).

\bibitem{Mele}
S.~M.~Young, S.~Zaheer, J.~C.~Y.~Teo, C.~L.~Kane, E.~J.~Mele, and A.~M.~Rappe,
Phys. Rev. Lett. {\bf 108}, 140405 (2012).

\bibitem{Manes:2011jk}
  J.~L.~Manes,
  %``Existence of bulk chiral fermions and crystal symmetry,''
  Phys. Rev. B {\bf 85}, 155118 (2012).
  %%[arXiv:1109.2581 [cond-mat.str-el]].
  %%CITATION = ARXIV:1109.2581;%%

\bibitem{Fang}
Z.~Wang, Y.~Sun, X.~Q.~Chen, C.~Franchini, G.~Xu, H.~Weng, X.~Dai, and Z.~Fang,
Phys. Rev. B {\bf 85}, 195320 (2012).
%arXiv:1202.5636

\bibitem{WangWeng}
Z.~Wang, H.~Weng, Q.~Wu, X.~Dai, and Z.~Fang,
Phys. Rev. B {\bf 88}, 125427 (2013).
%arXiv:1305.6780v2 [cond-mat.mtrl-sci].
%Three Dimensional Dirac Semimetal and Quantum Transports in Cd_3As_2

\bibitem{Borisenko}
S.~Borisenko, Q.~Gibson, D.~Evtushinsky, V.~Zabolotnyy, B.~Buchner, and R.~J.~Cava,
Phys. Rev. Lett. {\bf 113}, 027603 (2014).

\bibitem{Neupane}
M.~Neupane, S.-Y.~Xu, R.~Sankar, N.~Alidoust, G.~Bian, C.~Liu, I.~Belopolski, T.-R.~Chang,
H.-T.~Jeng, H.~Lin, A.~Bansil, F.~Chou, and M.~Z.~Hasan,
Nat. Commun. {\bf 5}, 3786 (2014).

\bibitem{Liu}
Z.~K.~Liu, B.~Zhou, Y.~Zhang, Z.~J.~Wang, H.~M.~Weng, D.~Prabhakaran, S.-K.~Mo, Z.~X.~Shen,
Z.~Fang, X.~Dai, Z.~Hussain, and Y.~L.~Chen,
Science {\bf 343}, 864 (2014).

\bibitem{Hook}
A.~A.~Burkov, M.~D.~Hook, and L.~Balents,
Phys. Rev. B {\bf 84}, 235126 (2011).

\bibitem{Turner}
A.~M.~Turner and A.~Vishwanath,
arXiv:1301.0330 [cond-mat.str-el].

\bibitem{Vafek}
O.~Vafek and A.~Vishwanath,
Ann. Rev. Cond. Mat. Phys. {\bf 5}, 83 (2014).

\bibitem{Savrasov}
X.~Wan, A.~M.~Turner, A.~Vishwanath, and S.~Y.~Savrasov,
Phys. Rev. B {\bf 83}, 205101 (2011).

\bibitem{Balents}
A.~A.~Burkov and L.~Balents,
Phys. Rev. Lett. {\bf 107}, 127205 (2011).

\bibitem{Cho}
G.~Y.~Cho, arXiv:1110.1939 [cond-mat.str-el].

\bibitem{1501.00060} H.~Weng, C.~Fang, Z.~Fang, A.~Bernevig, and X.~Dai,
  %``Weyl semimetal phase in noncentrosymmetric transition-metal monophosphides,''
Phys. Rev. X {\bf 5}, 011029 (2015).  
% arXiv:1501.00060 [cond-mat.mtrl-sci].

\bibitem{1501.00755} S.-M.~Huang, Su-Y.~Xu, I.~Belopolski, C.-C.~Lee, G.~Chang, B.~Wang, N.~Alidoust, G.~Bian, M.~Neupane, A.~Bansil, H.~Lin, and M.~Z.~Hasan,
  %``An inversion breaking Weyl semimetal state in the TaAs material class''
  arXiv:1501.00755 [cond-mat.mtrl-sci].

\bibitem{1502.00251}
C.~Zhang, Z.~Yuan, S.~Xu, Z.~Lin, B.~Tong, M.~Z.~Hasan, J.~Wang, C.~Zhang, and S.~Jia,
  %``Tantalum Monoarsenide: an Exotic Compensated Semimetal,"
arXiv:1502.00251 [cond-mat.mtrl-sci]

\bibitem{1502.03807}
S.-Y.~Xu, I.~Belopolski, N.~Alidoust, M.~Neupane, C.~Zhang, R.~Sankar, S.-M.~Huang, C.-C.~Lee, G.~Chang, B.~Wang, G.~Bian, H.~Zheng, D.~S.~Sanchez, F.~Chou, H.~Lin, S.~Jia, and M.~Z.~Hasan,
  %``Experimental realization of a topological Weyl semimetal phase with Fermi arc surface states in TaAs''
  arXiv:1502.03807 [cond-mat.mtrl-sci].

\bibitem{1502.04684}
B.~Q.~Lv, H.~M.~Weng, B.~B.~Fu, X.~P.~Wang, H.~Miao, J.~Ma, P.~Richard, X.~C.~Huang, L.~X.~Zhao, G.~F.~Chen, Z.~Fang, X.~Dai, T.~Qian, and H.~Ding,
  %``Discovery of Weyl semimetal TaAs"
arXiv:1502.04684 [cond-mat.mtrl-sci].
  %%CITATION = ARXIV:1502.04684;%%

\bibitem{1503.01304}
X.~Huang, L.~Zhao, Y.~Long, P.~Wang, D.~Chen, Z.~Yang, H.~Liang, M.~Xue, H.~Weng, Z.~Fang, X.~Dai, and G.~Chen,
  %``Observation of the chiral anomaly induced negative magneto-resistance in 3D Weyl semi-metal TaAs,''
  arXiv:1503.01304 [cond-mat.mtrl-sci].
  
\bibitem{Weyl-photonic} L.~Lu, Z.~Wang, D.~Ye, L.~Ran, L.~Fu, J.~D.~Joannopoulos, M.~Solja\v{c}i\'{c}, 
arXiv:1502.03438 [cond-mat.mtrl-sci].

\bibitem{Gorbar:2013qsa}
  E.~V.~Gorbar, V.~A.~Miransky, and I.~A.~Shovkovy,
  %``Engineering Weyl nodes in Dirac semimetals by a magnetic field,''
  Phys.\ Rev.\ B {\bf 88}, 165105 (2013).
  % [arXiv:1307.6230 [cond-mat.mes-hall]].
  %%CITATION = ARXIV:1307.6230;%%

\bibitem{Kim:2013dia}
H.-J.~Kim, K.-S.~Kim, J.~F.~Wang, M.~Sasaki, N.~Satoh, A.~Ohnishi, M.~Kitaura, M.~Yang, and L.~Li,
%``Dirac versus Weyl Fermions in Topological Insulators: Adler-Bell-Jackiw Anomaly in Transport Phenomena,''
 Phys. Rev. Lett. {\bf  111}, 246603 (2013).
%%[arXiv:1307.6990 [cond-mat.str-el]].
  %%CITATION = ARXIV:1307.6990;%%

\bibitem{Gorbar:2013dha}
  E.~V.~Gorbar, V.~A.~Miransky and I.~A.~Shovkovy,
  %``Chiral anomaly, dimensional reduction, and magnetoresistivity of Weyl and Dirac semimetals,''
Phys. Rev. B {\bf 89}, 085126 (2014).
%  Phys. Rev. B {\bf 89}, no. 8, 085126 (2014).
  %[arXiv:1312.0027 [cond-mat.mes-hall]].

 \bibitem{Son:2012bg}
  D.~T.~Son and B.~Z.~Spivak,
  %``Chiral Anomaly and Classical Negative Magnetoresistance of Weyl Metals,''
  Phys. Rev. B {\bf 88}, 104412 (2013).
  %[arXiv:1206.1627 [cond-mat.mes-hall]].

\bibitem{Nielsen:1983rb}
H.~B.~Nielsen and M.~Ninomiya,
%``Adler-bell-jackiw Anomaly And Weyl Fermions In Crystal,''
  Phys. Lett. B {\bf 130}, 389 (1983).
  %%CITATION = PHLTA,B130,389;%%


\bibitem{Haldane}
F.~D.~M.~Haldane, arXiv:1401.0529 [cond-mat.str-el].

\bibitem{Aji}
V.~Aji, Phys. Rev. B {\bf 85}, 241101 (2012).

\bibitem{Okugawa:2014ina}
  R.~Okugawa and S.~Murakami,
  %``Dispersion of Fermi arcs in Weyl semimetals and their evolutions to Dirac cones,''
  Phys. Rev. B {\bf 89}, 235315 (2014).
  %[arXiv:1402.7145 [cond-mat.mes-hall]].
  %%CITATION = ARXIV:1402.7145;%%

\bibitem{Hosur}
P.~Hosur, Phys. Rev. B {\bf 86}, 195102 (2012).

\bibitem{Vishwanath}
A.~C.~Potter, I.~Kimchi, and A.~Vishwanath,
Nature Commun. {\bf 5}, 5161 (2014).
%arXiv:1402.6342 (2014).
%Quantum Oscillations from Surface Fermi-Arcs in Weyl and Dirac Semi-Metals

\bibitem{Gorbar:2014qta}
  E.~V.~Gorbar, V.~A.~Miransky, I.~A.~Shovkovy, and P.~O.~Sukhachov,
  %``Quantum oscillations as a probe of interaction effects in Weyl semimetals in a magnetic field,''
  Phys. Rev. B {\bf 90}, 115131 (2014).
  %[arXiv:1407.1323 [cond-mat.str-el]].
  %%CITATION = ARXIV:1407.1323;%%

\bibitem{Gorbar:2014sja}
  E.~V.~Gorbar, V.~A.~Miransky, I.~A.~Shovkovy, and P.~O.~Sukhachov,
  %``Dirac semimetals $A_3Bi$ ($A$=Na,K,Rb) as $Z_2$ Weyl semimetals,''
  Phys. Rev. B {\bf 91}, 121101 (2015).
  %[arXiv:1412.5194 [cond-mat.str-el]].
  %%CITATION = ARXIV:1412.5194;%%

\bibitem{Yu}
W.~Zhang, R.~Yu, H.J.~Zhang, X.~Dai, and Z.~Fang,
New J. Phys. {\bf 12}, 065013 (2010).

\bibitem{Shen}
S.~Q.~Shen, {\sl Topological Insulators} (Springer, Heidelberg, 2012). 

\bibitem{science.1256742}
S.-Y. Xu, C. Liu, S. K. Kushwaha, R. Sankar, J. W. Krizan, I. Belopolski, M. Neupane, G. Bian, N. Alidoust, T.-R. Chang, H.-T. Jeng, C.-Y. Huang, W.-F. Tsai, H. Lin, P. P. Shibayev, F.-C. Chou, R. J. Cava, and M. Z. Hasan,
%``Observation of Fermi arc surface states in a topological metal,"%
Science {\bf 347}, 294 (2015).


\end{thebibliography}
\end{document}